\documentstyle[psfig]{mn}

\def\etal{{\it et al.\ }}
\def\eg{{\it e.g.\ }}

\def\ie{{\it i.e.\ }}

\def\spose#1{\hbox to 0pt{#1\hss}}
\def\approxlt{\mathrel{\spose{\lower 3pt\hbox{$\sim$}}
	\raise 2.0pt\hbox{$<$}}}
\def\approxgt{\mathrel{\spose{\lower 3pt\hbox{$\sim$}}
	\raise 2.0pt\hbox{$>$}}}
\def\approxpropto{\mathrel{\spose{\lower 3pt\hbox{$\sim$}}
	\raise 2.0pt\hbox{$\propto$}}}
\mathchardef\twiddle="2218

\def\multleft#1{\hbox to size{\vbox {\halign {\lft{##}\cr #1}}\hfill}\par}
\def\multright#1{\hbox to size{\vbox {\halign {\rt{##}\cr #1}}\hfill}\par}

\def\today{\ifcase\month\or January\or February\or March\or April\or May\or
      June\or July\or August\or September\or October\or November\or December\fi
      \space\number\day, \number\year}
\def\<{\thinspace}

\def\apc{\rm atom cm$^{-2}$}

\def\erg{{\rm\thinspace erg}}

\def\keV{{\rm\thinspace keV}}

\def\km{{\rm\thinspace km}}

\def\Mpc{{\rm\thinspace Mpc}}
\def\Msun{\hbox{$\rm\thinspace M_{\odot}$}}

\def\s{{\rm\thinspace s}}
\def\yr{{\rm\thinspace yr}}

\def\ergps{\hbox{$\erg\s^{-1}\,$}}

\def\kmps{\hbox{$\km\s^{-1}\,$}}

\def\Msunpyr{\hbox{$\Msun\yr^{-1}\,$}}

\def\kmpspMpc{\hbox{$\kmps\Mpc^{-1}$}}

\def\apc{\rm atom cm$^{-2}$}
\title[X-ray and gravitational lensing masses]
{Resolving the discrepancy between X-ray and gravitational lensing mass measurements
for clusters of galaxies}
\author[S.W. Allen]
{\parbox[]{6.in} {S.W. Allen \\
\footnotesize
Institute of Astronomy, Madingley Road, Cambridge CB3 OHA\\
}}
 
\begin{document}
\maketitle
\begin{abstract}

We present a detailed comparison of mass measurements 
for clusters of galaxies using ASCA and ROSAT X-ray data and constraints
from strong and weak gravitational lensing. Our results, for a sample of 
thirteen clusters (including six with massive cooling flows, five without 
cooling flows, and two intermediate systems) provide a consistent description 
of the distribution of gravitating matter in these systems.  
For the six cooling-flow clusters, which are the more dynamically-relaxed
systems, the X-ray 
and strong gravitational lensing mass measurements show excellent agreement. 
The core radii for the mass distributions are small,
with a mean value (using a simple isothermal parameterization) 
of $\sim 50h_{\rm 50}^{-1}$ kpc. These results imply that thermal pressure 
dominates over non-thermal processes in the support of 
the X-ray gas against gravity in the central regions of the cooling-flow
clusters, and that the hydrostatic assumption used in the X-ray mass 
determinations is valid.

For the non-cooling flow clusters, the masses determined from the strong
lensing data exceed the X-ray values by factors of $2-4$. 
However, significant offsets between the X-ray and lensing centres are observed,
indicating that the X-ray and strong-lensing data are probing different lines of sight
through the clusters. These offsets, and the generally complex dynamical states 
of the clusters inferred from their X-ray morphologies, lensing data 
and galaxy distributions, suggest that the 
gravitational potentials in the central regions of the non-cooling flow
systems are evolving rapidly, and that the assumption of hydrostatic equilibrium 
involved in the X-ray mass measurements is likely to have broken down. 
The discrepancies between the X-ray and strong lensing mass measurements  
may be reconciled if the dynamical activity has caused the 
X-ray analyses to overestimate the core radii of the dominant mass clumps in 
the clusters. Substructure and line-of-sight alignments of material towards 
the cluster cores may also contribute to the 
discrepancies since they will 
increase the probability of detecting gravitational arcs in the clusters 
and can enhance the lensing masses, without significantly
affecting the X-ray data.
On larger spatial scales, comparisons of the X-ray mass results with
measurements from weak gravitational lensing show 
excellent agreement for both cooling-flow  and non-cooling flow clusters.

Our method of analysis accounts for the effects of cooling flows 
on the X-ray data. We highlight the importance of this and show how the 
inappropriate use of simple isothermal models in the analysis of 
X-ray data for clusters with massive cooling flows will result in 
significant underestimates of their X-ray temperatures and masses.

\end{abstract}

\begin{keywords}
galaxies: clusters: general -- cooling flows -- intergalactic medium -- 
gravitational lensing -- X-rays: galaxies
\end{keywords}

\vskip -3in

\section{Introduction}

 Accurate measurements of the masses of clusters of galaxies provide a 
crucial observational constraint on cosmological models. 
Clusters are the largest gravitationally-bound objects known and 
represent rare peaks in the primordial density 
field on spatial scales of order 10 Mpc. The number densities and 
spatial distributions of clusters in a given mass range can be directly 
related to cosmological simulations and semi-analytic models 
(\eg Frenk \etal 1990; Evrard 1990; Henry \& Arnaud 1991; White, 
Efstathiou \& Frenk 1993; Viana \& Liddle 1996; Eke, Cole \& Frenk 1996;
Kitayama \& Suto 1996; Oukbir \& Blanchard 1997; Oukbir, Bartlett \& 
Blanchard 1997). Historically, measurements of the masses of clusters were 
made from optical studies of their galaxy populations. However, 
such studies are complicated by the presence of 
complex galaxy orbits, substructure (resulting from the growth of clusters 
through merger events) and projection effects (Lucey 1983; Sutherland 
1988; Frenk \etal 1990; van Haarlem, Frenk \& White 1997)  

 Currently, the two most promising techniques for obtaining accurate 
measurements of cluster masses are via X-ray observations and
observations of gravitational lensing by clusters. 
Clusters of galaxies are luminous X-ray sources, with typical 
luminosities ranging from a few $10^{43} - 10^{46}$ \ergps. 
The X-rays from clusters are primarily bremsstrahlung emission from the 
diffuse intracluster medium (ICM) that fills the deep gravitational 
potentials. The mass in X-ray gas exceeds the visible stellar 
mass by a factor of $1-5$, and typically contributes between 10 and 30 per cent of 
the total mass of the cluster (with the largest values observed for the 
most-massive systems; David, Jones \& Forman 1995; White \& Fabian 1995).
The X-ray emissivity is proportional to the square of the gas density and
accurately traces the three-dimensional cluster potentials. 
X-ray observations thus offer a method for identifying
clusters and determining 
cluster masses that is comparatively free from the projection affects
that complicate the optical studies (\eg Gioia \etal 1990; Ebeling \etal 1996,
1997). 

 Measurements of the masses of clusters from the X-ray data are based on the
assumption that the ICM is in hydrostatic equilibrium with the
gravitational potential of the cluster. The total mass profile is
determined once the radial profiles of the gas density and temperature are
known. The gas density profile can be accurately determined from X-ray
images. Measurements of the temperature profile, however, require 
detailed spatially-resolved spectroscopy. Although radial temperature profiles
have been determined for a few clusters (\eg Allen \& Fabian 1994; Nulsen
\& B\"ohringer 1995; Markevitch \& Vikhlinin 1997), 
in general the constraints are not firm and significant uncertainties remain,
particularly in the outer ($r > 1$ Mpc) regions of
clusters. For most systems only a mean emission-weighted X-ray
temperature, determined from an integrated cluster spectrum, is available.  
More precise information on the temperature profiles of clusters will become 
available in the near future, however, following
the launch of AXAF. 

  In contrast to the aforementioned optical 
galaxy dispersion and X-ray techniques, gravitational
lensing offers a method for 
measuring the projected surface density of matter through clusters that is
essentially free from assumptions about the dynamical state of the 
gravitating material (Fort \& Mellier 1994).
Recently, a number of studies have compared mass
measurements for clusters, using the X-ray and gravitational lensing
techniques. Miralda-Escud\'e and Babul (1995) presented an analysis
of the clusters Abell 1689 and Abell 2218 and noted a 
discrepancy of a factor $\sim 2$ between 
the X-ray and strong lensing mass determinations for these systems. 
Miralda-Escud\'e and Babul (1995) suggested a number of possible explanations 
for the observed discrepancy: (i) the clusters could have prolate ellipsoidal 
mass distributions. (ii) A superposition of mass clumps along 
the lines of sight through the systems. (iii) The X-ray gas 
could have a complex, non-isothermal temperature structure. 
(iv) Bulk and/or turbulent motions or magnetic fields could contribute 
significantly to the support of the ICM against gravity [this point was
also explored by Loeb \& Mao (1994) who suggested that 
non-thermal pressure support could completely account for the X-ray/strong 
lensing mass discrepancy in 
Abell 2218]. (v) The multiphase nature of the central ICM could result in
significant differences between emission-weighted and mass-weighted
temperatures for clusters, potentially biasing X-ray mass measurements to 
low values. The importance of this final point was discussed in detail
by Waxman \& Miralda-Escud\'e (1995), and will be further addressed in this 
paper. 

 Wu \& Fang (1997) compared mass estimates from 
optical galaxy dispersions, gravitational lensing and X-ray
methods for a large sample of data drawn from the literature. These authors 
concluded that the mass measurements from the galaxy dispersions and 
lensing data were generally in agreement, but were typically a factor $2-3$ 
larger than the X-ray-determined values. These authors suggested that 
non-thermal pressure support in the ICM, or simplifying assumptions
employed in their X-ray analysis, were likely to be responsible for the 
observed discrepancy. 

  Bartelmann \& Steinmetz (1996) used gas-dynamical simulations
to investigate the biases impinging on detections of strong gravitational
lensing by clusters of galaxies. These authors showed that clusters selected 
for their strong lensing properties are typically more dynamically active than
average clusters, with arcs occurring preferentially in clusters 
exhibiting substructure and non-equilibrium states. Bartelmann \&
Steinmetz (1996) concluded that in those clusters where the
X-ray-determined mass is not equal to the strong lensing mass, the
discrepancy is primarily due to enhancements of the lensing mass 
via projection effects. 

  Comparisons of weak lensing and X-ray mass measurements for clusters
provide a somewhat contrasting view to that obtained with the 
strong lensing data. Although few detailed comparisons have been made 
to date, in general such studies have inferred good 
agreement between the weak-lensing and X-ray determined masses; 
\eg the studies of Abell 2218 and Abell 2163 by Squires \etal (1996, 1997a).
Smail \etal (1997) present weak-lensing masses for twelve $z \sim 0.4$
clusters imaged with the Hubble Space Telescope (HST) and demonstrate 
reasonable agreement with the masses estimated 
from their X-ray luminosities and the empirical $L_{\rm
X}/T_{\rm X}$ relation. 

  X-ray observations of clusters of galaxies show that in the central
regions of most ($70-90$ per cent) clusters 
the cooling time of the ICM is significantly
less than the Hubble time (Edge \etal 1992; White, Jones \& Forman 1997; 
Peres \etal 1997).  The observed cooling leads to a slow net inflow of 
material towards the cluster centre; a process known as a 
cooling flow (Fabian 1994). The X-ray imaging data show that gas typically 
`cools out' throughout the central few tens to hundreds of kpc in the clusters, 
with ${\dot M}(r) \approxpropto r$, where ${\dot M}(r)$ is the integrated 
mass deposition rate within radius $r$. Recent spatially resolved X-ray 
spectroscopy has confirmed the presence of distributed cool 
(and rapidly cooling) gas in cooling flows, with a spatial distribution and 
luminosity in excellent agreement with the predictions from the imaging
data (Allen \& Fabian 1997). 
The natural state for a regular, relaxed cluster of galaxies 
appears to be with a cooling flow in its core. Once established, 
only a major merger event is likely to 
disrupt the central regions of a cluster to the extent that a cooling flow 
is `turned-off' (McGlynn \& Fabian 1984; Edge \etal 1992).

  In their combined X-ray and strong gravitational lensing study of the 
massive cooling-flow cluster PKS0745-191,  Allen, Fabian \& Kneib (1996a) 
demonstrated the importance of accounting for the effects of 
cooling flows on determinations of cluster masses from the X-ray data. 
These authors demonstrated excellent agreement between X-ray and strong 
lensing masses for PKS0745-191,  once the multiphase nature of the
X-ray emission from the cooling flow was accounted for. 
In contrast, when more simple single-phase
analyses of the X-ray data are employed (as has been the case in most 
previous studies) 
the total mass within the critical lensing radius can be underestimated
by as much as a factor $\sim 3$.  This point was further illustrated in the study of 
the more distant, massive cooling flow cluster Abell 1835 by Allen 
\etal (1996b). Both PKS0745-191 and Abell 1835 
appear dynamically relaxed at optical and X-ray wavelengths, in comparison
to many other famous lensing systems such as Abell 1689 and Abell 2218.  

  In this paper we present a detailed comparison of X-ray and lensing mass
measurements for 13 clusters of galaxies. Our sample 
includes 6 strong cooling flows, 5 non-cooling flows and 
2 intermediate systems (where the classifications are made according to
the fraction of the X-ray luminosity from the clusters contributed by
their cooling flows). We explore the relationships between the dynamical
states of the clusters (which relate to the presence or absence of 
cooling flows in these systems, as well as their morphological 
properties; Buote \& Tsai 1996b) and X-ray and gravitational-lensing 
measurements of their masses. We show how taking full account of the various
processes affecting the X-ray measurements can lead to a consistent 
picture for the distribution of gravitating matter in these systems. 
Throughout this paper, we assume $H_0$=50 \kmpspMpc, 
$\Omega = 1$ and $\Lambda = 0$.

\section{Observations and Data Reduction}
\begin{table*}
\vskip 0.2truein
\begin{center}
\caption{ASCA Observations}
\vskip 0.2truein
\begin{tabular}{ c c c c c c c }
\multicolumn{1}{c}{} &
\multicolumn{1}{c}{} &
\multicolumn{1}{c}{} &
\multicolumn{1}{c}{} &
\multicolumn{1}{c}{} &
\multicolumn{1}{c}{} &
\multicolumn{1}{c}{} \\
 \hline                                                             
Cluster         & ~ & Date        &   S0  &   S1  &  G2   &  G3    \\  
 \hline                                                             
&&&&&& \\                                                       
Abell 2744      & ~ & 1994 Jul 04 & 37605 & 26086 & 62749 & 62753  \\
PKS0745-191     & ~ & 1993 Nov 06 & 29146 & ----- & 37553 & 37553  \\
Abell 963       & ~ & 1993 Apr 22 & 29611 & 29039 & 29883 & 29881  \\
Abell 1689      & ~ & 1993 Jun 26 & 29575 & 23642 & 37817 & 37817  \\
RXJ1347.5-1145  & ~ & 1995 Jan 17 & 27882 & 17549 & 38968 & 38958  \\
MS1358.4+6245   & ~ & 1995 Apr 27 & 32532 & 30815 & 31981 & 31513  \\
Abell 1835      & ~ & 1994 Jul 20 & 34927 & 33976 & 33876 & 33870  \\
Abell 2163      & ~ & 1993 Aug 08 & 25126 & 18224 & 32760 & 32322  \\
Abell 2218      & ~ & 1993 Apr 30 & 28241 & 26054 & 37970 & 37968  \\
Abell 2219      & ~ & 1994 Aug 07 & 32705 & 31697 & 35849 & 35849  \\
MS2137.3-2353   & ~ & 1994 May 08 & 15167 & 15732 & 17035 & 17056  \\
Abell 2390      & ~ & 1994 Nov 13 & 6172  & 2632  & 10340 & 10338  \\
AC114           & ~ & 1995 Nov 09 & 36739 & 36295 & 35987 & 35971  \\
&&&&&& \\                                                                  
 \hline                                                             
&&&&&& \\                                                                  
\end{tabular}
\end{center}
\parbox {7in}
{Notes: A summary of the ASCA observations. Column 2 lists the date of 
observation. Columns $3-6$ list the effective exposure times (in seconds)
for the four ASCA detectors.
}
 \end{table*}

\begin{table*}
\vskip 0.2truein
\begin{center}
\caption{ROSAT HRI Observations}
\vskip 0.2truein
\begin{tabular}{ c c c c c r }
\multicolumn{1}{c}{} &
\multicolumn{1}{c}{} &
\multicolumn{1}{c}{} &
\multicolumn{1}{c}{} &
\multicolumn{1}{c}{} &
\multicolumn{1}{l}{} \\
 \hline                                                                        
Cluster         & ~ &  Date        & HRI    &    R.A. (J2000.)                           
&    ~~~Dec. (J2000.) \\  
 \hline                                                                        
&&&&& \\                                                                  
Abell 2744      & ~ & 1994 Dec 09  & 34256 &  $00^{\rm h}14^{\rm m}18.7^{\rm s}$ & $-30^{\circ}23'11''$ \\ 
PKS0745-191     & ~ & 1992 Oct 20  & 23750 &  $07^{\rm h}47^{\rm m}31.1^{\rm s}$ & $-19^{\circ}17'47''$ \\
Abell 963       & ~ & 1992 Nov 24  & 10104 &  $10^{\rm h}17^{\rm m}03.4^{\rm s}$ & $39^{\circ}02'51''$  \\
Abell 1689      & ~ & 1994 Jul 22/1995 Jun 24 & 22728 &  $13^{\rm h}11^{\rm m}29.1^{\rm s}$ & $-01^{\circ}20'40''$ \\
RXJ1347.5-1145  & ~ & 1995 Jan 28  & 15760 &  $13^{\rm h}47^{\rm m}31.0^{\rm s}$ & $-11^{\circ}45'11''$ \\
MS1358.4+6245   & ~ & 1993 May 14  & 15872 &  $13^{\rm h}59^{\rm m}50.8^{\rm s}$ & $62^{\circ}31'05''$  \\
Abell 1835      & ~ & 1993 Jan 22  & 2850  &  $14^{\rm h}01^{\rm m}02.0^{\rm s}$ & $02^{\circ}52'40''$  \\
Abell 2163      & ~ & 1994 Aug 13  & 36248 &  $16^{\rm h}15^{\rm m}45.9^{\rm s}$ & $-06^{\circ}08'58''$ \\
Abell 2218      & ~ & 1994 Jan 05 -- 1994 Jun 17  & 92856 &  $16^{\rm h}35^{\rm m}52.5^{\rm s}$ & $66^{\circ}12'29''$  \\
Abell 2219      & ~ & 1994 Jan 17  & 13242 &  $16^{\rm h}40^{\rm m}20.2^{\rm s}$ & $46^{\circ}42'29''$  \\
MS2137.3-2353   & ~ & 1994 Apr 24  & 13656 &  $21^{\rm h}40^{\rm m}15.2^{\rm s}$ & $-23^{\circ}39'41''$ \\
Abell 2390      & ~ & 1993 Nov 23  & 27764 &  $21^{\rm h}53^{\rm m}36.5^{\rm s}$ & $17^{\circ}41'45''$  \\
AC114           & ~ & 1993 May 17/1994 May 09  & 23192 &  $22^{\rm h}58^{\rm m}48.7^{\rm s}$ & $-34^{\circ}48'19''$ \\
&&&&& \\                                                                  
 \hline                                                                        
&&&&& \\                                                                  
\end{tabular}
\end{center}
\parbox {7in}
{Notes: A summary of the ROSAT HRI observations. Column 2 and 3  list the date 
of observation and exposure time (in seconds). Columns 4 and 5 list the 
coordinates of the centres of the X-ray emission from the clusters. 
}
\end{table*}

  Our sample consists of those clusters, reported in the
literature to exhibit strong gravitational lensing, for which at the
time of writing,  high-quality ASCA X-ray spectra and ROSAT High Resolution 
Imager (HRI) images, were available on the Goddard Space Flight Centre 
(GSFC) public archive (with the exception of the HRI image of
RXJ1347.5-1145, which was kindly provided by H. B\"ohringer \& S.
Schindler).

 The ASCA (Tanaka, Inoue \& Holt 1994) observations were made over a 
two-and-a-half year period between 1993 April
and 1995 November. The ASCA X-ray Telescope array (XRT) consists of four 
nested-foil telescopes, each focussed onto one of four detectors; two X-ray 
CCD cameras, the Solid-state Imaging Spectrometers (S0 and S1), and 
two Gas scintillation Imaging  Spectrometers (G2 and G3). The XRT 
provides a spatial resolution
of $\sim 3$ arcmin Half Power Diameter (HPD) in the energy range $0.3 - 12$
keV.  The SIS detectors provide excellent spectral resolution [$\Delta
E/E = 0.02(E/5.9 {\rm keV})^{-0.5}$] over a $22 \times 22$ arcmin$^2$ field of
view. The GIS detectors provide poorer energy resolution [$\Delta E/E =
0.08(E/5.9 {\rm keV})^{-0.5}$] over a larger circular field of view of 
$\sim 50$ arcmin diameter. Screened event lists were extracted from the ASCA
archive and were reduced using the FTOOLS package developed and 
supported by GSFC.  Standard reduction procedures, as recommended in the
GSFC ASCA Data Reduction Guide, were followed, including appropriate
grade selection,  gain  corrections and (where necessary)  manual
screening based on the individual instrument light curves. 

  The ROSAT HRI observations were carried out between 1991 November and 1995
June. The HRI provides a $\sim 5$ arcsec (FWHM) X-ray imaging 
facility (David \etal 1996). Reduction of the data was carried out with the Starlink 
ASTERIX package. X-ray images were extracted on a $2 \times 2$ arcsec$^2$ 
pixel scale, from which centres for the cluster X-ray emission 
were determined. Where more than one observation of a source was made, a
mosaicked image was constructed from the individual observations. 
For the cooling flow and intermediate clusters, 
the X-ray centres were identified with the peaks of the X-ray surface brightness 
distributions, which are easily determined from the HRI images. 
For the non-cooling flow clusters the X-ray emission is not as sharply-peaked
and for these systems we identify the X-ray centres with the results from
iterative determinations of the centroids of the emission
within a 2 arcmin radius of the cluster centres. (For Abell 2163 and AC114
a 1 arcmin radius aperture was better-suited and used. For Abell 2744, 2218 and 2219 
the use of either a 1 or 2 arcmin aperture does not significantly affect the
determinations of the X-ray centres). 

  We note here that the lensing cluster Abell 370 also has ASCA
and ROSAT HRI data available on the GSFC public archive but was not
included in our sample because the HRI data show 
that it is not a single, coherent structure but rather 
consists of a number of individual subclumps. 
The assumptions of spherical symmetry and hydrostatic equilibrium required for 
the X-ray mass modelling will therefore not apply. The X-ray images for
the other clusters included in the present sample do not exhibit any dramatic 
substructure that would clearly invalidate such assumptions. We note, however, the presence of 
an X-ray luminous subcluster, approximately 2.6 arcmin (850 kpc) to the northwest of Abell
2744 (AC118), visible in the HRI data. This subcluster is also
identified in the weak lensing analysis of Smail \etal (1997). 

 The details of the ASCA and ROSAT observations are summarized in Tables 1 
and 2, respectively. The basic X-ray properties of the target clusters 
are summarized in Table 3. 
 
\begin{table*}
\vskip 0.2truein
\begin{center}
\caption{X-ray properties of the cluster sample}
\vskip 0.2truein
\begin{tabular}{ c c c c c c c c }
\hline                                                                                                                               
\multicolumn{1}{c}{} &
\multicolumn{1}{c}{} &
\multicolumn{1}{c}{z} &
\multicolumn{1}{c}{$N_{\rm H}$} &
\multicolumn{1}{c}{$L_{\rm X,2-10}$} &
\multicolumn{1}{c}{$kT$} &
\multicolumn{1}{c}{${\dot M}_{\rm Spec}$} &
\multicolumn{1}{c}{$\Delta N_{\rm H}$} \\                            
                & ~ &        & ($10^{20}$ \apc) & ($10^{44}$ \ergps) & (keV)    & (\Msunpyr) & ($10^{20}$ \apc) \\
\hline                                                                                                                               
&&&&&&& \\
COOLING FLOWS &&&&&&& \\
&&&&&&& \\
PKS0745-191     & ~ &  0.103 & 42.4 & 29.5 & $8.7^{+1.6}_{-1.2}$   & $1460^{+350}_{-520}$  & $28^{+11}_{-13}$ \\  
RXJ1347.5-1145  & ~ &  0.451 & 4.9  & 93.5 & $26.4^{+7.8}_{-12.3}$ & $3480^{+340}_{-1150}$ & $27^{+30}_{-7}$ \\  
MS1358.4+6245   & ~ &  0.327 & 1.9  & 10.5 & $7.5^{+7.1}_{-1.5}$   & $690^{+350}_{-290}$   & $64^{+87}_{-38}$  \\  
Abell 1835      & ~ &  0.252 & 2.3  & 44.9 & $9.8^{+2.3}_{-1.3}$   & $1760^{+520}_{-590}$  & $32^{+16}_{-8}$ \\ 
MS2137.3-2353   & ~ &  0.313 & 3.6  & 16.6 & $5.2^{+1.8}_{-0.7}$   & $1470^{+880}_{-730}$  & $62^{+61}_{-28}$ \\  
Abell 2390      & ~ &  0.233 & 6.8  & 41.0 & $14.5^{+15.5}_{-5.2}$ & $1530^{+580}_{-1110}$ & $29^{+76}_{-15}$ \\  
&&&&&&& \\
INTERMEDIATE &&&&&&& \\
&&&&&&& \\
Abell 963       & ~ &  0.206 & 1.4  & 12.7 & $6.13^{+0.45}_{-0.30}$ & --- & --- \\  
Abell 1689      & ~ &  0.184 & 1.8  & 32.2 & $10.0^{+1.2}_{-0.8}$   & $350^{+290}_{-210}$ & $41^{+56}_{-18}$  \\  
&&&&&&& \\
NON COOLING FLOWS &&&&&&& \\
&&&&&&& \\
Abell 2744      & ~ & 0.308 & 1.6   & 30.9  & $7.75^{+0.59}_{-0.53}$  & --- & --- \\
Abell 2163      & ~ & 0.208 & 12.1  & 60.1  & $10.85^{+0.71}_{-0.63}$ & --- & --- \\
Abell 2218      & ~ & 0.175 & 3.2   & 10.8  & $7.18^{+0.50}_{-0.45}$  & --- & --- \\  
Abell 2219      & ~ & 0.228 & 1.8   & 38.0  & $9.46^{+0.63}_{-0.57}$  & --- & --- \\  
AC114           & ~ & 0.312 & 1.3   & 17.2  & $8.10^{+1.01}_{-0.85}$  & --- & --- \\  
&&&&&&& \\                                                                                                                         
\hline                                                                                                                               
\end{tabular}
\end{center}
\parbox {7in}
{Notes: Columns 2 and 3 list the cluster redshifts and Galactic column 
densities (from Dickey \& Lockman 1990). $L_{\rm X,2-10}$ values are the 
X-ray luminosities in the $2-10$ keV rest-frame of the source, 
determined from the G3 spectra. 
For the cooling flow and intermediate clusters, the temperatures
($kT$) were determined with a spectral model incorporating an 
intrinsically-absorbed, cooling flow component. For the non-cooling
systems, a more simple isothermal 
model with free-fitting absorption (assumed to lie at zero redshift) was
used. Column 6 lists the mass deposition rates from the cooling flows
determined from the ASCA spectra and
Column 7 the intrinsic absorbing column densities 
determined to act on the cooling flows. 
Errors bars are 90 per cent ($\Delta \chi^2 = 2.71$) confidence limits 
on a single interesting parameter.  Where no value for the mass
deposition rate is listed, this component was not statistically required by the data.
}
\end{table*}

\section{Analysis}

 The lensing clusters studied in this paper are drawn from a
larger sample of X-ray luminous systems discussed 
by Allen \etal (1997, in preparation). A more detailed description of the X-ray analysis
is included in that work, and only a brief summary is 
presented here. The method of X-ray analysis follows the multiphase technique 
employed in the studies of PKS0745-191 and Abell 1835 by Allen \etal
(1996a,b). A re-analysis of both of these clusters is included in the current work. 

 For the purposes of this paper, clusters are classified into three
categories; cooling flows, non-cooling flows and intermediate systems.
The cooling flows are those clusters with central cooling times 
$< 5 \times 10^{9}$ yr and for which the flux from cooling gas is spectrally 
determined to account for $\geq 20$ per cent of the total X-ray luminosity. 
Intermediate clusters are those systems with central cooling times $< 10^{10}$ yr 
and for which the cooling flows are spectrally
determined to contribute $< 20$ per cent of the total X-ray luminosity. 
Non-cooling flow systems are those clusters with central cooling times $>
10^{10}$ yr, and which show no spectral evidence
for cooling flow emission. For full details see Allen \etal (1997, in preparation).

\subsection{X-ray spectral analysis}

  Spectra were extracted from all four ASCA detectors (except for 
PKS0745-191, for which the S1 data were lost due to chip saturation problems)
in circular regions, centred on the X-ray centroids (Table 2). 
For the SIS data, the radii of the regions were 
adjusted to minimize the number of chip boundaries crossed 
(thereby minimizing the systematic
uncertainties introduced by such crossings) whilst covering
as large a region of the clusters as possible. Data from the regions between 
the chips were masked out and excluded. For the GIS data a constant 
extraction radius of 6 arcmin was used. 

  For the GIS observations, and SIS observations of clusters in 
regions of low Galactic column density
($N_{\rm H} \approxlt 5 \times 10^{20}$ \apc), background subtraction was 
carried out using the `blank sky' observations 
of high Galactic latitude fields complied during the 
performance verification stage of the ASCA
mission. For such data sets, the blank-sky
observations provide a reasonable
representation of the cosmic and instrumental backgrounds in the
detectors. The background data were screened and grade selected in the same 
manner as the target observations and background 
spectra were extracted over the same regions as the cluster spectra.  
For the SIS observations of clusters in directions of higher Galactic column 
density (PKS0745-191, Abell 2163, Abell 2390) background spectra were 
extracted from regions of the chips that were relatively free from 
foreground cluster emission.

  For the SIS data, response matrices were generated using the FTOOL SISRMG. 
Where the spectra covered more than one chip, response matrices were
created for each chip, which were then combined to form a
counts-weighted mean matrix. For the GIS analysis, the  response matrices 
issued by GSFC on 1995 March 6 were used.

  Modelling of the X-ray spectra was carried out using the XSPEC
spectral fitting package (version 9.0; Arnaud 1996). 
For the SIS data, only counts in pulse height analyser (PHA) channels
corresponding to energies between 0.6 and 10.0  \keV~  were included in the 
analysis (the energy range over which the calibration of the SIS 
instruments is best-understood). For the GIS data only counts in the energy 
range $1.0  - 10.0$ \keV~were used. 
The spectra were grouped before fitting to ensure a minimum of 20  
counts per PHA channel, allowing $\chi^2$ statistics to be used.

 The spectra have been modelled using the plasma codes of Kaastra \& Mewe
(1993; incorporating the Fe L calculations by Liedhal in XSPEC version 9.0) 
and the photoelectric absorption models of Balucinska-Church \& McCammon 
(1992). The spectra were examined with
a series of models. For clarity, in this paper we only report those
relevant results from the best-fitting models. (A more complete 
discussion is given by Allen \etal 1997, in preparation). 
The data from all four ASCA detectors were simultaneously fitted, 
with the parameters forced to take the same values across the data
sets. The exceptions to this were the emission measures of the cluster gas 
in the different detectors which, due to the different 
extraction radii used (and residual uncertainties in the flux calibration
of the instruments), were allowed to fit independently. 

 The non-cooling flow clusters were found to be well-described by a simple 
isothermal plasma model, where the temperature,
metallicity, absorbing column density and emission measures were included
as free parameters in the fits. The best-fit column densities for 
the non-cooling flow clusters were generally found to be 
in reasonable agreement with the Galactic values although were,
typically, slightly higher (Allen \etal 1997, in 
preparation; see also Section 4.2). In contrast, the cooling-flow clusters 
required the introduction of a second
emission component with a lower mean
temperature, which was also required to be intrinsically absorbed. For 
consistency with the imaging analysis presented in Section 3.2 we have 
modelled this cooler component as a constant-pressure cooling flow, in
which gas is assumed to cool from the ambient cluster temperature, 
following the prescription of Johnstone \etal (1992). (This is model is 
referred to as spectral model C by
Allen \etal 1997, in preparation). The 
plasma code of Kaastra \& Mewe (1993; incorporating the Fe L calculations
by Liedhal) was again used. 

 Column 5 of Table 3 lists the best-fit temperatures (for the ambient
cluster gas) and 90 per cent ($\Delta \chi^2 = 2.71$) confidence limits for the clusters. 
The spectrally-determined mass deposition rates and intrinsic X-ray
absorbing column densities are listed in Columns 6 and 7 of
that Table.

\subsection{X-ray imaging analysis and mass results}

 The analysis of the HRI imaging data was carried out using an 
updated version of the deprojection code of Fabian \etal (1981; see 
also White \etal 1997 for details). 
Azimuthally-averaged X-ray surface brightness profiles 
were determined for each cluster from the HRI images. 
The profiles were background-subtracted, 
corrected for telescope vignetting and re-binned 
to provide sufficient counts in each radial bin 
for the deprojection analysis to be successfully carried out
(bin sizes of 8-24 arcsec were used).

With the X-ray surface brightness profiles as the primary input, 
and under assumptions of spherical symmetry and hydrostatic equilibrium 
in the ICM, the deprojection technique can be used 
to study the basic properties of the intracluster gas 
(temperature, density, pressure, cooling rate) as a 
function of radius. The code uses a monte-carlo method to determine the 
statistical uncertainties on the results and 
incorporates the latest HRI spectral response matrix issued by GSFC. 
The metallicity and absorbing column density of the cluster gas 
were fixed at the values determined from the spectral analysis in Section 3.1 

The deprojection code requires the total 
mass profiles for the clusters (which define the
pressure profiles) to be specified. We have iteratively determined 
the mass profiles that result in deprojected
temperature profiles (which approximate the mass-weighted
temperature profiles in the clusters) that are isothermal within the regions probed by
the HRI data (the central $0.5-1$ Mpc) and which are consistent with the spectrally-determined
temperatures from Section 3.1. The assumption of 
approximately isothermal mass-weighted temperature profiles in the central regions 
of the clusters is supported by the following evidence: firstly, 
ASCA observations of nearby cooling flows show that 
in the central regions of these systems the gas is multiphase, but 
that the bulk of the X-ray gas there has a temperature close
to the cluster mean  (\eg Fukazawa \etal 1994, Ohashi \etal 1997, Fabian
\etal, in preparation). Secondly, the combined X-ray and
gravitational lensing studies of the cooling-flow clusters PKS0745-191
and Abell 1835 (Allen \etal 1996a,b) demonstrated that approximately isothermal
mass-weighted temperature profiles are required to consistently 
explain the X-ray imaging and spectral data for these systems, and result
in good agreement between the X-ray and gravitational-lensing masses 
for the clusters.  Thirdly, the use of approximately constant 
mass-weighted temperature profiles implies a more plausible range of initial
density inhomogeneities in the clusters than would be the case if the 
temperature profiles decreased within the clusters cores 
(Thomas, Fabian \& Nulsen 1987).  Finally, the use of approximately isothermal 
mass-weighted temperature profiles in the deprojection analyses 
leads to independent 
determinations of the mass deposition profiles from the cooling flows, 
from the X-ray spectra and imaging data, in excellent 
agreement with each other (Allen \& Fabian 1997). 
We note that the assumption of a constant
mass-weighted deprojected temperature profile is
consistent with measurements of a decreasing emission-weighted 
temperatures in the cores of many cooling-flow clusters
(Waxman \& Miralda-Escud\'e 1995). 

The mass profiles for the clusters were parameterized as isothermal spheres 
(Equation 4-125 of Binney \& Tremaine 
1987) with adjustable core radii, $r_{\rm c}$, and velocity dispersions, 
$\sigma$. The core radii were adjusted until the temperature profiles 
determined from the deprojection code became isothermal. The velocity 
dispersions were then adjusted  until the temperatures determined from the deprojection code
came into agreement with the spectrally-determined values. 
Errors on the velocity dispersions are the range of
values that result in isothermal deprojected temperature profiles 
that are consistent, at the 90 per cent confidence limit, with the 
spectrally-determined temperatures. Errors on the core radii 
denote the range of values that are consistent with isothermality 
in the deprojected temperature profile. Errors on the core radii are only listed
for the cooling-flow and intermediate clusters since, for the non-cooling
flow systems, the large core radii inferred are likely to be 
due to recent merger events having disrupted the central regions of the
clusters and invalidated the assumption of hydrostatic equilibrium 
(Sections 4.4, 4.6). [We note that an initial estimate for 
the pressure in the outermost radial bin used in the analysis is also 
required by the deprojection code. These values were also 
determined iteratively, under the assumption of isothermality. 
The uncertainties on these
pressure estimates do not significantly affect the results presented here.
Note also that the core radii determined from the deprojection analysis are 
similar, though not identical, to the  values determined from simple 
`$\beta-$model' fits to the X-ray surface brightness profiles 
\eg Jones \& Forman 1984.]. 
The mass distributions determined from the deprojection analysis 
are summarized in Columns 5 and 6 of Table 4.  In Column 7 we list
the projected masses, inferred from these distributions, within the
critical radii defined by 
the gravitational arcs in the clusters (Section 3.3). The mass distributions 
are assumed to extend to radii of 3 Mpc.  

  Finally, we note that although the deprojection method of Fabian \etal 
(1981) is essentially a single-phase technique, it produces results in 
good agreement with more detailed multiphase treatments 
(Thomas, Fabian \& Nulsen 1987) and, due to its simple applicability at 
large radii in clusters, is better-suited to the 
present project. Detailed results on the 
cooling flows in these clusters, also determined from the deprojection 
code, are presented by Allen \etal (1997, in preparation).

\subsection{Strong gravitational lensing analysis}

The lensing data used in this
paper have been drawn from the literature and are summarized in Table 4 
(Columns 2 and 3, with references listed in the caption). 
The X-ray modelling presented in Sections 3.1 and 3.2 was carried out
under the assumption of spherical symmetry in the 
cluster mass distributions. The use of a spherically-symmetric geometry in the 
X-ray analyses, where the underlying cluster mass distributions
are ellipsoidal, will tend to slightly 
overestimate the X-ray gas pressure and, therefore, gravitating mass as a 
function of radius. However, the use of spherical models is not unreasonable since,
the X-ray gas (in hydrostatic equilibrium) will trace the cluster
potentials, which will be more spherical than the mass distributions, 
particularly at large radii. The spherical analysis also avoids
degeneracies associated with the unknown oblate/prolate nature of the mass
distributions. The X-ray masses determined with the spherical modelling 
should be accurate to $\sim 10$ per cent (\eg Buote \& Tsai 1996a).

In the first case, for simplicity and to be consistent 
with the X-ray analysis, we have carried out a basic lensing analysis 
using circularly-symmetric models for the lensing potentials. As will 
be shown in Section 4.3, the use of 
more realistic, elliptical mass models can reduce the masses 
within the arc radii by up to 40 per cent, although values of $\sim 20$
per cent are more typical (see also Bartelmann 1995).
However, such corrections are not significant in 
comparison to the factor $\approxgt 2$ discrepancies between the X-ray and 
lensing masses reported for clusters like Abell 1689 and 2218 
(Miralda-Escud\'e \& Babul 1995).

  For a circular mass distribution the projected mass within the tangential
critical radius, assumed to be equal to the arc radius, 
$r_{\rm arc}$, is given by

\begin{equation}
M_{\rm arc}(r_{\rm arc}) ~ =
\frac{c^2 }{4 G} \left( \frac{D_{\rm arc}} {D_{\rm clus} D_{{\rm
arc-clus}}}
\right) ~ r_{\rm arc}^2
\end{equation}

 where $D_{\rm clus}$, $D_{\rm arc}$ and $D_{\rm arc-clus}$ are respectively 
the angular diameter distances from the observer to the cluster, the observer 
to the lensed object, and the  cluster to the lensed object.  (We note that
where the lensed features do not lie exactly on the critical curves,
small overestimates of the lensing masses are likely to result). 
Where redshifts for the arcs are not available,  we calculate 
masses for assumed arc redshifts of 1.00 and 2.00.
The masses within the arc radii, determined from the lensing analysis, are 
summarized in Table 4 (Column 4).

\section{Results and Discussion}

\subsection{The ratio of strong-lensing to X-ray masses}

 \begin{table*}
\vskip 0.2truein
\begin{center}
\caption{Strong lensing and X-ray mass measurements} 
\vskip 0.2truein
\begin{tabular}{ c c l l l c c c c l l }
\hline
\multicolumn{1}{c}{} &
\multicolumn{1}{c}{} &
\multicolumn{3}{c}{LENSING} &
\multicolumn{1}{c}{} &
\multicolumn{3}{c}{X-RAY} &
\multicolumn{1}{c}{} &
\multicolumn{1}{c}{RATIO} \\
&&&&&&& \\
             & ~ & $r_{\rm arc}$ & $z_{\rm arc}$   &  $M_{\rm arc}$     & ~ &   $\sigma$ & $r_{\rm c}$ &  $M_{\rm X}$        & ~ &   ~~~~~~~$M_{\rm arc}$/$M_{\rm X}$        \\
             & ~ &  (kpc)        &                 &  ($10^{13}$ \Msun) & ~ & (\kmps) & (kpc) & ($10^{13}$ \Msun)          & ~ &                                \\
\hline                                                                                                                               
&&&&&&& \\
COOLING FLOWS &&&&&&& \\
&&&&&&& \\
PKS0745-191    & ~ & 45.9      & 0.433       &  2.99              & ~ & $930^{+90}_{-70}$   & 37.5 ($\pm 5$) & $3.16^{+0.64}_{-0.46}$ & ~ & $0.95^{+0.16}_{-0.16}$    \\
RXJ1347.5-1145 & ~ & 240       & 1.00 (2.00) &  51.0 (35.8)       & ~ & $1850^{+270}_{-500}$ & 75 ($\pm 15$) & $68.1^{+21.4}_{-31.8}$  & ~ & $0.75^{+0.65}_{-0.18}$  ($0.53^{+0.46}_{-0.13}$)         \\
MS1358.4+6245  & ~ & 121       & 4.92        &  8.27              & ~ & $830^{+340}_{-80}$   & 40 ($\pm 30$) & $7.03^{+6.97}_{-1.29}$  & ~ & $1.18^{+0.26}_{-0.59}$          \\
Abell 1835     & ~ & 150       & 1.00 (2.00) &  18.1 (15.4)       & ~ & $1000^{+120}_{-70}$  & 50 ($\pm 20$) & $12.6^{+3.2}_{-1.7}$    & ~ & $1.44^{+0.22}_{-0.29}$ ($1.22^{+0.19}_{-0.25}$)       \\
Abell 2390     & ~ & 174       & 0.913       &  25.4              & ~ & $1190^{+520}_{-240}$ & 60 ($\pm 20$) & $21.2^{+22.6}_{-7.7}$   & ~ & $1.20^{+0.68}_{-0.62}$         \\
MS2137.3-2353  & ~ & 88.0      & 1.00 (2.00) &  6.15 (4.98)       & ~ & $830^{+140}_{-50}$   & 32 ($\pm 25$) & $5.19^{+1.88}_{-0.62}$  & ~ & $1.18^{+0.17}_{-0.31}$ ($0.96^{+0.13}_{-0.26}$)            \\
&&&&&&& \\
&&&&&&& \\
INTERMEDIATE &&&&&&& \\
&&&&&&& \\
Abell 963      & ~ &   79.8    & 0.771       &   5.85             & ~ & $750^{+50}_{-50}$ & 80 ($\pm 25$) & $3.29^{+0.47}_{-0.41}$    & ~ & $1.78^{+0.25}_{-0.22}$         \\
Abell 1689     & ~ &   183     & 1.00 (2.00) &   29.5 (26.4)      & ~ & $990^{+60}_{-50}$ & 80 ($\pm 15$) & $15.3^{+2.0}_{-1.5}$      & ~ &  $1.93^{+0.21}_{-0.22}$ ($1.73^{+0.18}_{-0.20}$)      \\
&&&&&&& \\
&&&&&&& \\
NON COOLING-FLOWS &&&&&&& \\
&&&&&&& \\
Abell 2744(2)   & ~ &  119.6   & 1.00 (2.00) & 11.36 (9.23)       & ~ & $930^{+60}_{-50}$   & 450 & $2.78^{+0.37}_{-0.29}$ & ~ & $4.09^{+0.47}_{-0.48}$ ($3.32^{+0.39}_{-0.39}$)      \\
Abell 2163      & ~ &   67.7   & 0.73        & 4.29               & ~ & $1050^{+50}_{-50}$  & 300 & $1.74^{+0.17}_{-0.16}$ & ~ & $2.47^{+0.25}_{-0.22}$     \\
Abell 2218($\#359$)   & ~ &  79.4    & 0.702       & 6.23         & ~ & $830^{+30}_{-40}$   & 230 & $1.89^{+0.14}_{-0.18}$ & ~ & $3.30^{+0.34}_{-0.23}$         \\
Abell 2218($\#384$)   & ~ &  84.8    & 2.515       & 5.70         & ~ & $830^{+30}_{-40}$   & 230 & $2.14^{+0.16}_{-0.20}$ & ~ & $2.66^{+0.28}_{-0.18}$         \\
$^*$Abell 2219N & ~ &  79.3    & 1.00 (2.00) & 5.17 (4.48)        & ~ & $950^{+50}_{-70}$   & 250 & $2.59^{+0.25}_{-0.32}$ & ~ & $2.00^{+0.28}_{-0.18}$ ($1.73^{+0.24}_{-0.15}$)       \\
$^*$Abell 2219L & ~ &  110.2   & 1.00 (2.00) & 9.98 (8.65)        & ~ & $950^{+50}_{-70}$   & 250 & $4.53^{+0.45}_{-0.60}$ & ~ & $2.20^{+0.34}_{-0.20}$ ($1.91^{+0.29}_{-0.17}$)           \\
AC114 (S1+D1/S2+D2)   & ~ &  67.6    & 1.86        & 2.98         & ~ & $910^{+80}_{-70}$   & 300 & $1.30^{+0.24}_{-0.19}$ & ~ & $2.29^{+0.39}_{-0.35}$ \\
&&&&&&& \\
\hline
&&&&&&& \\
\end{tabular}
\end{center}
\parbox {7in}
{Notes: The masses within the arc radii determined from the 
spherically-symmetric strong-lensing analyses (Section 3.3) and 
X-ray modelling (Sections 3.1, 3.2). Errors on the X-ray masses and the
strong lensing to X-ray mass ratios are 90 per cent confidence limits.
No statistical error has been associated with the lensing masses. (For
an estimate of the systematic uncertainties associated with the 
lensing results see the discussion of the analyses with the more detailed lensing
models; Section 4.3). References for the lensing data are as follows:
details for PKS0745-191 from Allen \etal (1996). 
RXJ1347.7-1145 arc radius from Schindler \etal (1997).  MS1358.4+6245 arc 
radius and redshift from Franx \etal (1997). Abell 1835 arc radius
from Edge \etal (in preparation). 
MS2137-2353 arc radius from Fort \etal (1992). 
Abell 2390 arc radius and redshift from Pell\'o \etal (1991). 
Lensing details for Abell 1689 and Abell 2163 from
Miralda-Escud\'e \& Babul (1995) and references therein. Abell 963 arc radius 
from Lavery \& Henry (1988) and redshift from Ellis \etal (1991).
Abell 2744 (AC118) arc radius from Smail \etal (1991).
Arc radii for Abell 2218 from Kneib \etal (1995). Redshift for arc
$\#359$ from Pell\'o \etal 1992. Redshift for arc $\#384$ from Ebbels
\etal (1996). Arc radii for Abell 2219 from Smail \etal (1995a).
Arc radius and redshift for AC114 from Smail \etal
(1995b). $^*$ The X-ray-determined mass for Abell 2219 includes a $3 \times 10^{12}$
\Msun~singular isothermal sphere truncated at 30 kpc, which improved 
the isothermality of the deprojected temperature profile. 
}
\end{table*}

\begin{figure*}
\centerline{\hspace{3cm}\psfig{figure=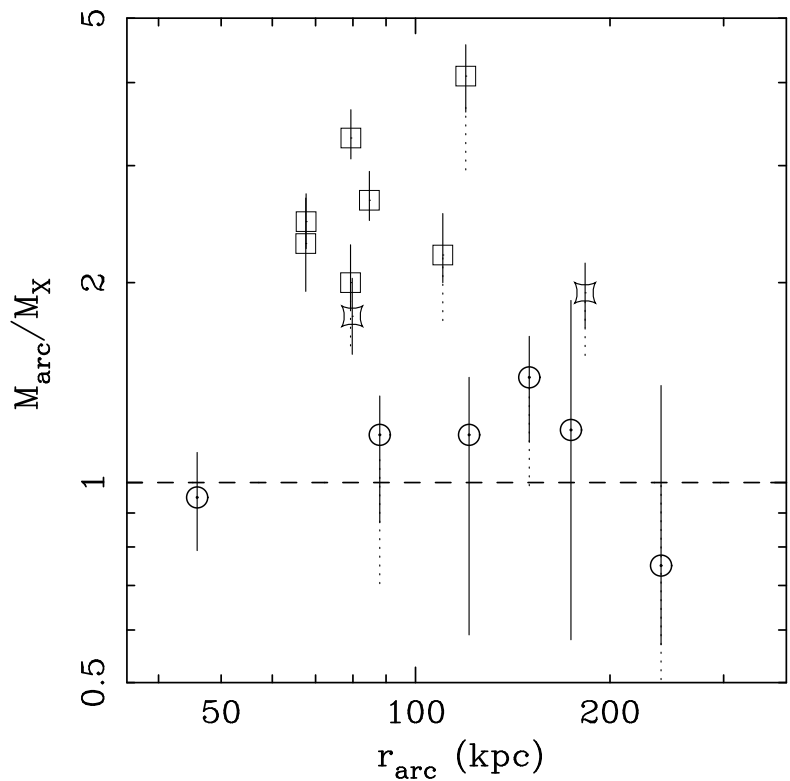,width=0.9\textwidth
,angle=270}}
\caption{The ratio of the strong lensing to X-ray mass measurements as
a function of the arc radius (as detailed in Table 4). The measurements for the cooling-flow clusters 
are plotted as circles. The non-cooling systems are plotted as squares. The two
intermediate systems (Abell 963 and Abell 1689) are indicated by pinched
squares. The solid error bars denote the 90 per cent confidence limits 
on the mass ratios, given the redshift of the arc (where known) or 
for an assumed arc redshift, $z_{\rm arc} = 1.00$.
Dotted error bars (where plotted)  denote the confidence 
limits for an assumed arc redshift of $z_{\rm arc} =2.00$. 
For the cooling-flow clusters, good agreement between the strong lensing and
X-ray mass measurements is observed. For the non-cooling flow 
systems,  the lensing mass exceeds the X-ray mass by 
a factor of $2-4$. For the two intermediate clusters the lensing mass 
exceeds the X-ray determined value by a smaller factor of $1.7-2.0$. 
The origins of the discrepancies between the strong lensing and X-ray
masses for the non-cooling flow clusters are explored in Section 4.
}
\end{figure*}

The ratios of the mass measurements from the strong lensing (Section 3.3)
and X-ray (Sections 3.1, 3.2) data are listed in Table 4 (Column 8). 
Fig. 1 shows the ratios of the lensing and X-ray masses as a function of 
arc radius. The results demonstrate a clear segregation between the
cooling flow and non-cooling flow systems. 

 All of the cooling flow clusters in our sample exhibit good agreement between their X-ray 
and strong-lensing masses. In particular, for those clusters with measured arc 
redshifts, excellent agreement between the X-ray and strong lensing masses is 
observed. For the non-cooling flows, the strong lensing masses 
exceed the X-ray masses by factors of $2-4$. For the two 
intermediate systems, where the detections of cooling flows from the
X-ray data are more marginal, the lensing masses are again 
enhanced with respect to the X-ray values, although by a smaller factor 
($1.7-2.0$). We find excellent agreement 
with the results of Miralda Escud\'e and Babul (1995) for the 
three (non-cooling flow) clusters in common with that study; 
Abell 1689, 2163 and 2218. 

 Our results support the conclusions drawn by Allen \etal (1996a,b) that 
thermal pressure dominates over magnetic pressure, turbulence and bulk 
motions in the central regions of the 
relaxed cooling-flow clusters, and that the hydrostatic
assumption adopted in the X-ray analysis of such systems is valid. 
In all cases where discrepancies between the X-ray and strong 
lensing masses occur, the clusters appear dynamically active 
and have either small (in comparison to their total X-ray luminosities) 
or no cooling flows. Clusters like Abell 1689, AC114, Abell 2163 and Abell 2218 have 
unusually high velocity dispersions (Gudehus 1989, Couch \& Sharples 1987, 
Squires \etal 1997a, Le Borgne, Pell\'o \& Sanahuja 1992), given their 
X-ray luminosities, and exhibit clear substructure in their X-ray emission, galaxy 
distributions and total matter distributions (see also Section 4.3 and references
therein). Merger events will tend to complicate the temperature structure 
in clusters, and generate turbulent and bulk motions which may contribute 
to the support of the X-ray gas against gravity. 
Although the discrepancies between the X-ray and strong lensing masses 
resulting from such processes should not, in general, exceed 50 per cent 
(Navarro, Frenk \& White 1995; 
Schindler 1996; Evrard, Metzler \& Navarro 1996; 
Roettiger, Burns \& Loken 1996) at smaller radii (comparable to the cluster
core radii) their effects may be more important (see Section 4.6). 

Substructure and line-of-sight alignments of material towards the 
cluster cores are also likely to contribute to the 
mass discrepancies since they will 
increase the probability of detecting gravitational arcs in the 
clusters, and enhance the masses determined from the lensing data 
(Bartelmann \& Steinmetz 1996). However,  
since magnetic fields are expected to be stronger in the cores of 
cooling-flow, rather than non-cooling flow systems 
(\eg Soker \& Sarazin 1990), magnetic pressure seems unlikely to contribute 
significantly to the differences between the strong lensing and X-ray 
masses observed.

\subsection{The effects of cooling flows on the X-ray data}
\begin{table*}
\vskip 0.2truein
\begin{center}
\caption{The effects of cooling flows on the X-ray masses}
\vskip 0.2truein
\begin{tabular}{ c c c c c c }
\multicolumn{1}{c}{} &
\multicolumn{1}{c}{} &
\multicolumn{1}{c}{} &
\multicolumn{1}{c}{} &
\multicolumn{1}{l}{} &
\multicolumn{1}{l}{} \\
\hline
              & ~ &  $\Delta kT$           &  $M_{\rm X}$            & $M_{\rm X, I}$       &     RATIO \\
              & ~ &     (keV)              &  ($10^{13}$ \Msun)      & ($10^{13}$ \Msun)      & ($M_{\rm X}$/$M_{\rm X, I}$)   \\
\hline
&&&&& \\
COOLING FLOWS &&&&& \\
&&&&& \\
PKS0745-191    & ~ & $2.2^{+1.8}_{-1.4}$   & $3.16^{+0.64}_{-0.46}$  & $2.34^{+0.11}_{-0.12}$ &  $1.35^{+0.36}_{-0.25}$  \\
RXJ1347.5-1145 & ~ & $16.0^{+8.6}_{-13.2}$ & $68.1^{+21.4}_{-31.8}$  & $27.0^{+2.1}_{-2.1}$   &  $2.52^{+0.79}_{-1.27}$  \\
MS1358.4+6245  & ~ & $1.8^{+7.7}_{-2.3}$   & $7.03^{+6.97}_{-1.29}$  & $5.44^{+0.77}_{-0.58}$ &  $1.29^{+1.59}_{-0.37}$  \\
Abell 1835     & ~ & $2.5^{+2.6}_{-1.7}$   & $12.6^{+3.2}_{-1.7}$    & $9.13^{+0.65}_{-0.43}$ &  $1.38^{+0.44}_{-0.27}$  \\
MS2137.3-2353  & ~ & $0.8^{+2.2}_{-1.2}$   & $5.19^{+1.88}_{-0.62}$  & $4.46^{+0.35}_{-0.46}$ &  $1.16^{+0.61}_{-0.21}$  \\
Abell 2390     & ~ & $5.6^{+16.5}_{-6.5}$  & $21.2^{+22.6}_{-7.7}$   & $13.1^{+1.7}_{-1.6}$   &  $1.62^{+2.19}_{-0.71}$  \\
&&&&& \\		                            
INTERMEDIATE &&&&& \\		                            
&&&&& \\		                            
Abell 963      & ~ &  ---                  &  $3.29^{+0.47}_{-0.41}$ & $3.29^{+0.27}_{-0.17}$ &  $1.00^{+0.21}_{-0.19}$ \\
Abell 1689     & ~ &  $0.8^{+1.6}_{-1.2}$  & $15.3^{+2.0}_{-1.5}$    & $14.0^{+0.8}_{-0.6}$   &  $1.10^{+0.19}_{-0.17}$ \\
&&&&& \\		      
\hline		      
&&&&& \\
\end{tabular}
\end{center}
\parbox {7in}
{Notes: The differences between the temperatures and masses determined
from the X-ray analyses with the cooling-flow and
isothermal spectral models. The errors on the masses 
($M_{\rm X}$ and $M_{\rm X, I}$ respectively) are 90 per cent confidence
limits. The errors on the temperature differences 
($\Delta kT$) and mass ratios ($M_{\rm X}$/$M_{\rm X, I}$) mark the
maximum and minimum values consistent with the joint 90 per cent confidence
limits on the results obtained with cooling flow and isothermal models. 
}
\end{table*}

\begin{figure}
\centerline{\hspace{2.5cm}\psfig{figure=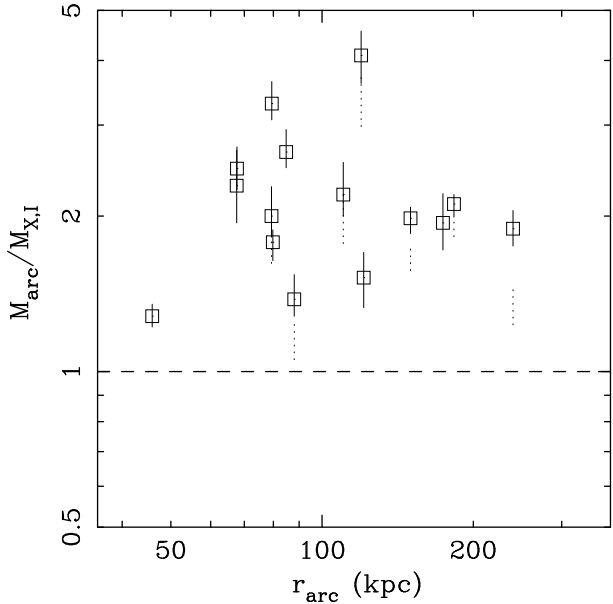,width=0.7\textwidth
,angle=270}}
\caption{The ratio of the strong lensing to X-ray masses 
as a function of the arc radius (determined with the 
spherically-symmetric lensing models) when no 
account for the effects of cooling flows on the X-ray data is made. 
Simple, isothermal X-ray spectral models have been used for all clusters,
including the cooling-flow systems. This figure (and comparison to Fig. 1)
illustrates how the failure to 
account for the presence of cooling
flows in the X-ray analyses can lead to the false conclusion that 
the strong lensing mass invariably exceeds the X-ray determined mass,
even in the relaxed, cooling-flow systems.
}
\end{figure}

The results presented in Section 4.1 demonstrate excellent agreement between 
the X-ray and strong-lensing masses for the 
cooling-flow clusters in our sample. It is crucial to note, however, 
that such agreement would not have been obtained if 
the multiphase nature of the X-ray emission from these clusters 
were not accounted for in the X-ray modelling. 

 The analysis of the ASCA data for the cooling flow 
(and intermediate) clusters in Section 3.1 incorporated a cooling-flow 
component (with intrinsic absorption) in the spectral modelling. 
The cooling-flow component accounts for 
the spectral signature from material cooling out of the X-ray waveband 
from the ambient cluster temperature. The luminosity of this component,
which is a free parameter in the fits, is generally 
found to be in excellent agreement with the cooling rates
determined independently from the deprojection analysis of the imaging data
(Allen \etal 1997, in preparation), lending strong support to the 
validity of this model. If the cooling flow component were
not incorporated into the spectral analysis, and a more simple isothermal
spectral model were inappropriately used instead, the measured 
temperatures would significantly underestimate the `true' values 
listed in Table 3. 

 The differences between the temperatures determined using the models
incorporating the cooling-flow components and the results obtained using the
more-simple isothermal spectral models are summarized in Table 5. 
Also listed in that Table are the total masses within the arc radii 
determined with the cooling-flow ($M_{\rm X}$; as in Table 4) and isothermal 
($M_{\rm X, I}$) spectral models, and the ratios of these 
values ($M_{\rm X}$/$M_{\rm X, I}$). We see that the failure to account 
for the complexities in the X-ray spectra
due to the presence of cooling flows will typically result in 
underestimates of the cluster masses by $10-40$ per cent. 
For RXJ1347.5-1145, the hottest and most 
X-ray luminous cluster in our sample, the underestimation is even more
severe; a factor $2- 3$. This is primarily due to the difficulty in 
constraining the temperatures of the hottest clusters 
($kT \approxgt 10$ keV) using the $0.6-10.0$ keV bandpass of ASCA. 
Such effects completely account for the discrepancy between
the X-ray and strong lensing masses for RXJ1347.5-1145 reported
by Schindler \etal (1997). [We note that exact agreement between the
strong lensing and X-ray masses for RXJ1347.5-1145 is achieved 
for an X-ray temperature of $kT \sim 19 (13.5)$ keV, for an assumed 
redshift for the arc of 1.00 (2.00), which is consistent with our
spectral constraints.] The two intermediate clusters in our sample, 
Abell 963 and Abell 1689, have lower fractions of their X-ray luminosities 
contributed by their cooling flows (Allen \etal 1997,  in preparation) and 
the errors incurred by not accounting for the effects of the cooling flows 
on their X-ray determined masses are therefore less severe.

 These results are shown in graphical form in Fig. 2., where we plot the
ratios of the strong lensing and X-ray masses, determined with the
isothermal models, as a function of the arc radius. We see that for all clusters
$M_{\rm arc} > M_{\rm X, I}$ . (Note that for the non-cooling flow clusters 
$M_{\rm X, I} = M_{\rm X}$ in  Table 4). This figure, together with Fig. 1,
clearly illustrates how the failure to account for the presence of cooling
flows in the X-ray analyses can lead to the incorrect conclusion that 
the strong lensing mass invariably exceeds the X-ray determined mass.

Finally, in this Section, we note that our 
analysis of the non-cooling flow clusters was carried out 
with the absorbing column density included as as a free parameter in the
spectral fits. 
Column densities in excess of the Galactic values were statistically
required for Abell 2744, Abell 2163, Abell 2219 and AC114 (Allen \etal 1997, in preparation). 
With the column densities fixed at nominal Galactic values
(Dickey \& Lockman 1990) the measured temperatures were somewhat higher; 
$kT = 11.0^{+0.8}_{-0.7}$,  $13.8^{+0.8}_{-0.7}$, $12.4^{+0.8}_{-0.7}$ and
$9.8^{+1.0}_{-0.9}$ keV for Abell 2744, Abell 2163, Abell 2219 and AC114, respectively. 
The corresponding 
X-ray masses within the arc radii are then $4.1 \times 10^{13}$ for Abell 2744, 
$2.3 \times 10^{13}$ for Abell 2163, $3.1 \times 10^{13}$ and $5.5
\times 10^{13}$ \Msun~for the N and L arcs of Abell 2219, and
$1.7 \times 10^{13}$ for AC114.  The ratios of the strong lensing to X-ray 
masses are reduced to  2.8 (2.2) for Abell 2744, 1.9 for 
Abell 2163,  1.7 (1.4) and 1.8 (1.6) for the N and L arcs in Abell 2219,
and 1.8 for AC114. Although the temperature results are therefore 
sensitive, in detail, to the 
modelling of the absorbing column density, the differences are 
not enough to account for the discrepancies
between the strong lensing and X-ray mass measurements for the non-cooling 
flow clusters.

\subsection{More detailed lensing models} 

Several of the clusters discussed in this paper are well-studied 
lensing systems. From their detailed study of Abell 2218, Kneib \etal (1995) 
demonstrated that the observed arc(let) configuration implies 
a mass distribution consisting of two clumps centred on the two brightest galaxies. 
The orientation and ellipticities of the projected potentials due to 
these clumps trace the external isophotes of brightest ellipticals. 
[Similar results on the orientation of the dark matter
potentials relative to the isophotes of the dominant cluster galaxies 
were obtained from the lensing studies of 
MS2137.3-2353 and Abell 370 by Mellier, Fort \& Kneib (1993)
and Kneib \etal (1993). These results may be compared to 
similar findings based on the galaxian and X-ray gas distributions in other clusters 
\eg Porter, Schneider \& Hoessel (1991), Allen \etal (1995).]  
The lensing results on Abell 2218, and comparison 
to a ROSAT HRI X-ray image of the cluster (from a 
shorter 11.5 ks exposure) lead  Kneib \etal (1995) to suggest that 
Abell 2218 is undergoing a subcluster merger event, which may 
have shocked the central X-ray gas and caused it to deviate from 
hydrostatic equilibrium. From their modelling, Kneib
\etal (1995) determine a mass within the ellipse traced by arc
$\#384$ of $6.1 \times 10^{13}$ \Msun, similar to the value 
of $5.7 \times 10^{13}$ \Msun~inferred from the simple spherical model.
Miralda-Escud\'e \& Babul (1995) showed that the application of a
multi-component lensing model, which can account for the positions of the brightest
arcs in Abell 2218, predicts a lensing mass within the circular aperture 
defined by arc $\#359$ of $\sim 87$ per cent of the value inferred from the
simple spherical model. This implies $M_{\rm arc}/M_{\rm X} = 2.87^{+0.30}_{-0.20}$, 
which is again similar to the result listed in Table 4.
(Recall that the errors quoted on the $M_{\rm arc}/M_{\rm X}$ values 
denote the largest and smallest values consistent with the 
joint confidence limits on the X-ray
masses and lensing results).

From their detailed lensing analysis of the cooling-flow cluster 
MS2137.3-2353, Mellier \etal
(1993) determined a mass within the external critical radius of 
$M_{\rm arc} = 3-7 \times 10^{13}$ \Msun. This is consistent with the
value determined from the circularly-symmetric model and is
in excellent agreement with the X-ray-determined mass measurement of 
$M_{\rm X} = 5.2^{+1.9}_{-0.6} \times 10^{13}$ \Msun~(implying 
$M_{\rm arc}/M_{\rm X} = 0.96^{+0.56}_{-0.54}$).
Mellier \etal (1993) also concluded that the core radius of
the lensing potential in MS2137.3-2353 is small ($\sim 50$ kpc),  
in good agreement with the X-ray result reported here (see also 
Section 4.6).

 For Abell 1689, Miralda-Escud\'e \& Babul (1995) showed that the
application of a mass model consisting of (at least) two clumps is 
required to reproduce the positions of the brightest arcs. 
The mass within the circular aperture defined by the dominant 
arc, determined with their model, is slightly larger (104 per cent) 
than the value inferred from the  simple spherical model, and gives 
$M_{\rm arc}/M_{\rm X} = 2.00^{+0.22}_{-0.23}$ ($1.80^{+0.19}_{-0.21}$). 

  Smail \etal (1995a) present results from a detailed lensing analysis of
Abell 2219 using a multi-component mass model. These authors determine a projected 
mass within 100 kpc of the cluster centre of $1.1 \pm 0.2 \times
10^{14}$ \Msun, in reasonable agreement (although slightly larger than) 
the values suggested by the simple spherically-symmetric model
(Table 4). The X-ray-determined mass within this projected radius 
is $3.54^{+0.38}_{-0.51} \times 10^{13}$ \Msun, implying a
strong-lensing/X-ray mass ratio of 
$M_{\rm arc}/M_{\rm X} = 3.11^{+1.18}_{-0.81}$. 
Smail \etal (1995a) also infer a core radius for 
the dominant mass clump of $46 \pm 20$ kpc.

  Allen \etal (1996a) present results from a more detailed 
lensing analysis of PKS0745-191.  These authors show that the 
application of an elliptical potential leads to a projected mass 
within the arc radius of $2.5^{+0.2}_{-0.4} \times 10^{13}$ \Msun, 
approximately 20 per cent lower than the value inferred from the spherical
model. The implied $M_{\rm arc}/M_{\rm X}$ value is then 
$0.79^{+0.21}_{-0.22}$. The lensing data for PKS0745-191 
also suggest a small core radius of $\sim 40$ kpc, in good agreement with
the value inferred from the X-ray  analysis.

 Pierre \etal (1996) present a more sophisticated lensing analysis of the cooling-flow
cluster Abell 2390 (employing a two-component lensing model). These
authors determine a projected mass within the arc radius of 
$1.6 \pm 0.2 \times 10^{14}$ \Msun. This
value is $\sim 40$ per cent lower than the value determined from the
simple, circular lensing model, but is in good agreement with the X-ray
measurement of $M_{\rm X} = 2.1^{+2.3}_{-0.8} \times 10^{14}$
\Msun~reported here, and implies $M_{\rm arc}/M_{\rm X} =
0.76^{+0.62}_{-0.44}$.  Pierre \etal (1996) adopt a core radius 
for the mass distribution of the dominant clump in their lensing model of
60 kpc, identical to that inferred from the X-ray analysis.

 Natarajan \etal (1997) present results from a detailed 
analysis of lensing data for AC114.
These authors determine masses within radii of 75, 150 and 500 kpc of the
cluster centre of $4.2 \pm 0.1 \times 10^{13}$,  $1.20 \pm 0.15 \times
10^{14}$, and $4.0 \pm 0.4 \times 10^{14}$ \Msun, respectively. These
measurements compare to X-ray determined values within the same radii of 
$1.59^{+0.30}_{-0.23} \times 10^{13}$, $5.82^{+1.07}_{-0.86} \times
10^{13}$, and $3.40^{+0.63}_{-0.50} \times 10^{14}$ \Msun. 
This implied lensing/X-ray mass ratios at these radii are then
$2.64^{+0.52}_{-0.47}$, $2.06^{+0.66}_{-0.56}$, and 
$1.18^{+0.34}_{-0.29}$, respectively. 
The Natarajan \etal (1997) value at 75 kpc is
similar to that listed in Table 4. 

 In conclusion, we see that although the more detailed lensing analyses 
refine the results on the strong-lensing/X-ray mass ratios, the
conclusions drawn from Section 4.1,  using the simple
spherically-symmetric lensing models, remain essentially unchanged. 
The $M_{\rm arc}/M_{\rm X}$ ratios for cooling-flow clusters 
(the more dynamically-relaxed systems) show excellent agreement. 
For the non-cooling flow clusters, the masses inferred from the 
strong lensing data are $\sim 2-4$ times larger than the X-ray
measurements. We shall now explore the reasons for this discrepancy.

\subsection{Offsets between the lensing and X-ray centres in non-cooling flow 
clusters}

\begin{table*}
\vskip 0.2truein
\begin{center}
\caption{Offsets between the X-ray and lensing centroids}
\vskip 0.2truein
\begin{tabular}{ c c c r c c c c c }
 \hline                                                                        
\multicolumn{1}{c}{} &
\multicolumn{1}{c}{} &
\multicolumn{2}{c}{Lensing Centre} &
\multicolumn{1}{c}{} &
\multicolumn{2}{c}{Offset} &
\multicolumn{1}{c}{} &
\multicolumn{1}{c}{$r_{\rm arc}$}  \\
                & ~ & R.A. (J2000.)                      &   ~~~Dec. (J2000.)   & ~ & (arcsec) & (kpc)  & ~ & (kpc) \\  
 \hline                                                                        
&&&&&&&& \\                                                                  
COOLING FLOWS &&&&&&&& \\                                                                  
&&&&&&&& \\                                                                  
PKS0745-191     & ~ & $07^{\rm h}47^{\rm m}31.3^{\rm s}$ & $-19^{\circ}17'41''$ & ~ & 6.6  & 16.6 & ~ & 45.9 \\
RXJ1347.5-1145  & ~ & $13^{\rm h}47^{\rm m}30.7^{\rm s}$ & $-11^{\circ}45'11''$ & ~ & 4.4  & 30.0 & ~ & 240  \\
MS1358.4+6245   & ~ & $13^{\rm h}59^{\rm m}50.7^{\rm s}$ & $62^{\circ}31'05''$  & ~ & 0.7  & 4.0  & ~ & 121  \\
Abell 1835      & ~ & $14^{\rm h}01^{\rm m}02.2^{\rm s}$ & $02^{\circ}52'40''$  & ~ & 3.0  & 14.8 & ~ & 150  \\
MS2137.3-2353   & ~ & $21^{\rm h}40^{\rm m}15.3^{\rm s}$ & $-23^{\circ}39'41''$ & ~ & 1.4  & 7.9  & ~ & 88.0 \\
Abell 2390      & ~ & $21^{\rm h}53^{\rm m}36.9^{\rm s}$ & $17^{\circ}41'43''$  & ~ & 6.1  & 28.6 & ~ & 174  \\
&&&&&&&& \\                                                                  			        
INTERMEDIATE &&&&&&&& \\                                                                  	        
&&&&&&&& \\                                                                  			        
Abell 963       & ~ & $10^{\rm h}17^{\rm m}03.8^{\rm s}$ & $39^{\circ}02'47''$  & ~ & 6.1  & 26.3 & ~ & 79.8  \\
Abell 1689      & ~ & $13^{\rm h}11^{\rm m}29.6^{\rm s}$ & $-01^{\circ}20'29''$ & ~ & 13.3 & 52.9 & ~ & 183   \\
&&&&&&&& \\                                                                  			        
NON COOLING FLOWS &&&&&&&& \\                                                                  	        
&&&&&&&& \\                                                                  			        
Abell 2744      & ~ & $00^{\rm h}14^{\rm m}20.8^{\rm s}$ & $-30^{\circ}24'03''$ & ~ & 58.7 & 328  & ~ & 119.6   \\ 
Abell 2163      & ~ & $16^{\rm h}15^{\rm m}49.1^{\rm s}$ & $-06^{\circ}08'43''$ & ~ & 50.0 & 217  & ~ & 67.7   \\
Abell 2218      & ~ & $16^{\rm h}35^{\rm m}49.5^{\rm s}$ & $66^{\circ}12'43''$  & ~ & 22.9 & 87.8 & ~ & 79.4/84.8 \\
Abell 2219      & ~ & $16^{\rm h}40^{\rm m}19.8^{\rm s}$ & $46^{\circ}42'41''$  & ~ & 12.7 & 58.7 & ~ & 79.3/110.2 \\
AC114           & ~ & $22^{\rm h}58^{\rm m}48.4^{\rm s}$ & $-34^{\circ}48'10''$ & ~ & 9.7 & 54.6  & ~ & 67.6  \\
&&&&&&&& \\                                                                  
 \hline                                                                        
&&&&&&&& \\                                                                  
\end{tabular}
\end{center}
\parbox {7in}
{Notes: The lensing centres and offsets (in arcsec and kpc) with respect
to the X-ray centres listed in Table 2. For comparison, the arc radii, within
which the strong-lensing masses are evaluated, are also listed. For
Abell 2218 and 2219 the radii for both of the two brightest arcs
are given.  
}
\end{table*}

 The X-ray emission from a cooling-flow cluster is typically 
regular (with an approximately ellipsoidal symmetry) 
and sharply-peaked on to a position co-incident with,
or close to, to the position of the optically-dominant cluster galaxy 
(\eg Allen \etal 1995, Allen \etal 1996b).
In such clusters, the X-ray gas and arc(let)s 
should trace the same cluster potential. 
In non-cooling flow systems, however, the situation is less
clear. Clusters without cooling flows generally exhibit more
substructure in their X-ray images (Buote \& Tsai 1996b) and appear
more complex in their galaxian and dark matter distributions. Such
clusters are often inferred to be undergoing major subcluster merger events. 
Non-cooling flow clusters do not have a sharply-defined peak to their X-ray
emission in the same manner that cooling flow clusters do ({\it c.f.}
Section 2). Thus, although detailed lensing studies show that the arc(let)
configurations in such clusters are still typically centred on 
the dominant cluster galaxies (non-cooling flow clusters often have more
than a single dominant galaxy) these galaxies are not 
necessarily coincident with the centres of the 
X-ray emission from the clusters. In such circumstances, simple 
comparisons of X-ray and strong lensing mass measurements, such as those 
presented in Sections $4.1-4.3$, may not be applicable.

 We have examined the alignment of the X-ray centroids (Table 2) 
with the centroids of the matter distributions inferred from are the
lensing studies (which are defined as the 
optical centres of the dominant cluster galaxies around which the
arc(let)s are observed). The galaxy positions were measured from 
the Space Telescope Science Institute Digitized Sky Survey (hereafter DSS).
In Table 6 we list the DSS coordinates for the relevant galaxies and the 
separations (in arcsec and kpc) between the lensing and X-ray centroids.
The accuracy of the optical and X-ray coordinates are such that 
offsets between these positions of $\approxgt 10$ arcsec should be considered 
significant ({\it c.f.} David \etal 1996).

 The results listed in Table 6 show that for all of the cooling flow
clusters the X-ray and lensing centres are consistent with each other, in
agreement with previous results (\eg Allen \etal 1995).  
For non-cooling flow systems, however, significant offsets between
the X-ray and lensing centres are observed. 
In all cases, the size of these offsets are comparable to or larger than the 
radii at which the arcs are observed. 
We note that an offset of $\sim 20$ arcsec between 
between the X-ray centroid and the dominant 
galaxy in Abell 2218 was previously noted by Markevitch (1997). 
Our results show that the X-ray and lensing mass measurements discussed
in Sections $4.1-4.3$ are actually probing different lines of sight
through the clusters and, therefore, that direct comparisons 
of these values are not strictly valid. 

  Within the context of stable, spherically-symmetric mass models,  
the projected mass through a region of fixed radius, centred on any
position other than the cluster centre, will always be less than the mass 
through the centre. Thus the discrepancies between the X-ray and lensing
masses are not immediately explained by the offsets between the X-ray and
lensing centres. However, the results on the offsets show 
that the X-ray gas in the central regions of the non-cooling flow clusters
is not in hydrostatic equilibrium  with the gravitational potentials 
inferred from the lensing data. In Section 4.6 we shall show how the
breakdown of the hydrostatic assumption can lead the X-ray 
measurements to significantly underestimate the true cluster masses.

\subsection{A comparison with weak lensing results} 

\begin{figure*}
\centerline{\hspace{3cm}\psfig{figure=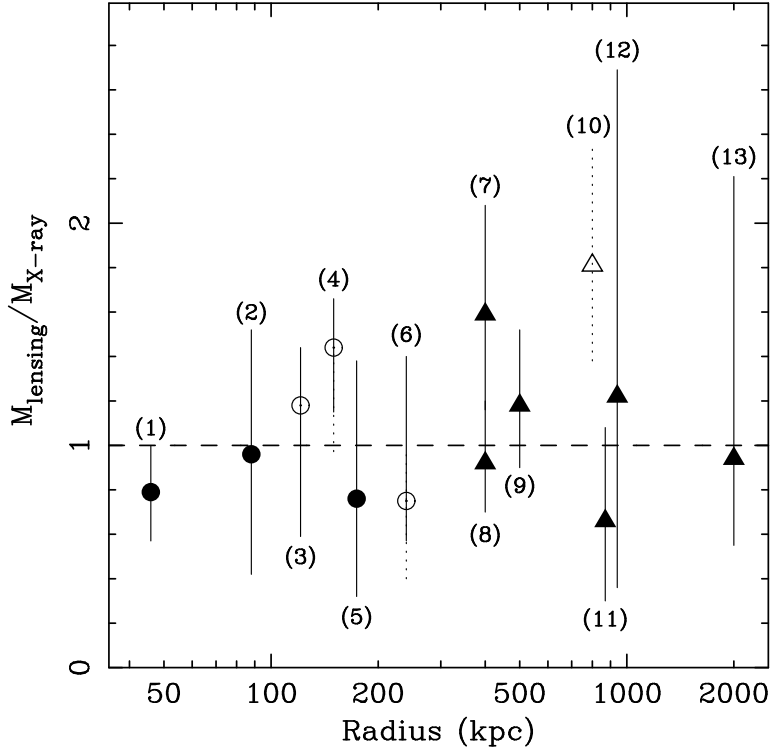,width=0.9\textwidth
,angle=270}}
\caption{The ratio of lensing (strong and weak) and X-ray 
masses for those clusters for which a reliable and direct comparison of
these values can be made. The ratios of the strong-lensing and X-ray masses 
are plotted as circles. Strong lensing results are only presented for the 
cooling-flow clusters since, for the
non-cooling flow and intermediate systems, the hydrostatic assumption is
not expected to hold (Sections 4.4, 4.6). Filled circles show the results
obtained with the detailed lensing models (Section 4.3) 
and open circles the results from the simple, spherically-symmetric lensing
models (which are only used when results from more-detailed modelling 
are not available).
The weak-lensing results (Section 4.5) are plotted as triangles. The open
triangle is the Squires \etal (1996) result for Abell 2218 at
a radius 800 kpc. The figure demonstrates excellent agreement
between the gravitational lensing and X-ray masses for all clusters
in our sample, for which reliable comparisons of these values can be made. 
The numbers in parentheses are used to identify the clusters; (1)
PKS0745-191, (2) MS2137.3-2353, (3) MS1358.4+6245, (4) Abell 1835, (5)
(5) Abell 2390, (6) RXJ1347.5-1145, (7) Abell 2744 , (8) Abell 2218 at
400 kpc, (9) AC114, (10) Abell 2218 at 800kpc, (11) Abell 2163, 
(12) Abell 2390, (13) RXJ1347.5-1145.}
\end{figure*}

\begin{table*}
\vskip 0.2truein
\begin{center}
\caption{Weak lensing and X-ray mass measurements}
\vskip 0.2truein
\begin{tabular}{ c c c c c c }
\multicolumn{1}{c}{} &
\multicolumn{1}{c}{} &
\multicolumn{1}{c}{} &
\multicolumn{1}{l}{} &
\multicolumn{1}{l}{} \\
\hline
               & ~ &  $r_{\rm weak}$ &  $M_{\rm weak}$    & $M_{\rm X, weak}$   &     RATIO \\
               & ~ &     (kpc)       &  ($10^{13}$ \Msun) & ($10^{13}$ \Msun)   & ($M_{\rm weak}$/$M_{\rm X, weak}$)   \\
\hline
&&&&& \\
COOLING FLOWS &&&&& \\
&&&&& \\
RXJ1347.5-1145 & ~ & 2000 & $340 \pm 80$   & $360^{+110}_{-170}$  &  $0.94^{+1.27}_{-0.39}$  \\
Abell 2390     & ~ & 940  & $100 \pm 40$   & $82^{+87}_{-30}$     &  $1.22^{+1.47}_{-0.86}$  \\
&&&&& \\
NON COOLING FLOWS &&&&& \\
&&&&& \\
Abell 2744     & ~ & 400  & $37.0 \pm 6.4$ & $23.3^{+3.1}_{-2.4}$ &  $1.59^{+0.49}_{-0.43}$  \\
Abell 2163     & ~ & 870  & $50 \pm 25$    & $76.3^{+7.5}_{-7.0}$ &  $0.66^{+0.42}_{-0.36}$  \\
Abell 2218(a)  & ~ & 800  & $78 \pm 14$    & $43.1^{+3.2}_{-4.0}$ &  $1.81^{+0.54}_{-0.43}$  \\
Abell 2218(b)  & ~ & 400  & $21.0 \pm 3.8$ & $22.9^{+1.7}_{-2.2}$ &  $0.92^{+0.28}_{-0.22}$  \\
AC114          & ~ & 500  & $40 \pm 4$     & $34.0^{+6.2}_{-5.0}$ &  $1.18^{+0.34}_{-0.28}$  \\
 &&&&& \\		      
\hline		      
&&&&& \\
\end{tabular}
\end{center}
\parbox {7in}
{Notes: A summary of the weak lensing mass measurements ($M_{\rm weak}$)
and X-ray determined masses within the same radii ($M_{\rm X, weak}$).
Column 2 lists the radii ($r_{\rm weak}$) within which the weak-lensing mass
measurements were made. Errors on the $M_{\rm X, weak}$ values are 90 per
cent confidence limits. Errors on the mass ratios 
mark the maximum and minimum values consistent
with the joint confidence limits on the $M_{\rm weak}$  and $M_{\rm X,
weak}$ values. References for the weak 
lensing data are as follows: RXJ1347.5-1145 from Fischer \& Tyson (1997). 
Abell 2390 from Squires \etal (1997b).  Abell 2744 (AC118) from 
Smail \etal (1997). Abell 2163 from Squires \etal (1997a). 
Abell 2218(a) from 
Squires \etal (1996) and (b) from Smail \etal (1997). 
AC114 from Natarajan \etal (1997).
}
\end{table*}

   A number of the clusters discussed in this paper have also been the
subject of detailed weak-lensing analyses. Studies of weak lensing by
clusters of galaxies probe the projected matter distributions 
on spatial scales $\sim 1$ Mpc, significantly larger than the 
offsets between the X-ray and lensing centres 
determined for the non-cooling flow clusters (Section 4.4).

  From their study of Abell 2218, 
Squires \etal (1996) determine a lower bound to the mass within an 
800 kpc (3.5 arcmin) radius of the cluster centre of $7.8 \pm 1.4 \times 
10^{14}$ \Msun. This compares to an X-ray-determined mass within the
same region of $4.31^{+0.32}_{-0.40} \times 10^{14}$ \Msun. 
The Squires \etal (1996) weak-lensing mass 
thus exceeds the X-ray determined value for the central 800 kpc 
by a factor $\sim 1.8$.  Within a smaller 400 kpc (radius) region, however, 
Squires \etal (1996) determine a mass of $2.4 \pm 0.6 \times 10^{14}$
\Msun~(this value has been estimated from their Fig. 16) in good agreement
with the X-ray measurement of $2.29^{+0.17}_{-0.22} \times 10^{14}$ \Msun~from 
the analysis presented here. 
Smail \etal (1997) also present results from a weak-lensing study of 
HST images of Abell 2218 from which they measure 
a mass within the central 400 kpc (radius) region of the cluster of 
$2.10 \pm 0.38 \times 10^{14}$ \Msun, in excellent agreement with the 
Squires \etal (1996) and X-ray results. 
The weak-lensing and X-ray results for Abell 2218 thus 
suggest an unusual (non-isothermal)
projected mass profile between radii of
400 and 800 kpc [with a density gradient flatter than $r^{-1}$. 
We note that the weak-lensing mass 
for Abell 2218 within 800 kpc (Squires \etal 1996) appears high
given that this mass is comparable to the value determined
for Abell 2390 within a similar aperture (see below), despite the 
fact that the $2-10$ keV X-ray luminosity of Abell 2390 is 
$\sim 4$ times higher than that of 
Abell 2218. Such conclusions are not significantly affected 
by the remodelling of the X-ray mass profiles with 
the smaller core radii, discussed in Section 4.6.]

  From their weak-lensing study of Abell 2163, Squires \etal (1997a)
determine (from their Fig. 5) a mass within a 200
arcsec (870 kpc) radius aperture of $\sim 5 \times 10^{14}$
\Msun~(with a factor $\sim 2$ uncertainty), in 
good agreement with the X-ray determined value, $M_{\rm X, weak} = 
7.6^{+0.8}_{-0.7} \times 10^{14}$ \Msun, reported here. 
In addition, from their study of the cooling-flow cluster Abell 2390, 
Squires \etal (1997b) measure (from their Fig. 3) a mass 
within $r = 200$ arcsec (940 kpc) of $10 \pm 4 \times 10^{14}$
\Msun, in excellent agreement with the X-ray determined value of 
$8.2^{+8.7}_{-3.0} \times 10^{14}$ \Msun. 

 From their weak lensing analysis of HST images, Smail \etal (1997) 
measure a mass within the central 400 kpc (radius) region of
Abell 2744 (AC118) of $3.70 \pm 0.64 \times 10^{14}$ \Msun.
This compares to the value determined from the X-ray analysis 
presented here of $2.33^{+0.31}_{-0.24} \times 10^{14}$ \Msun~ 
(implying a weak-lensing/X-ray mass ratio of $1.59^{+0.49}_{-0.43}$).

 From their analysis of weak lensing in the most X-ray luminous cluster of
galaxies known, RXJ1347.5-1145, Fischer \& Tyson (1997) determine a mass
within 2 Mpc of $3.4 \pm 0.8 \times 10^{15}$ \Msun. This is in excellent
agreement with the value of $3.6^{+1.1}_{-1.7} \times 10^{15}$ \Msun~ 
determined from the multiphase X-ray analysis presented here.

  The weak-lensing masses for the clusters in our sample, 
and the X-ray-determined masses within the same
regions, are summarized in Table 7. The values generally
exhibit good agreement, for both cooling flow and non-cooling flow
systems. We note again, however, that such agreement 
would not have been obtained for the cooling-flow
clusters if the effects of the cooling flows on the X-ray 
data had not been accounted for (Section 4.2). 
We conclude that the enhancements of the lensing masses with
respect to the X-ray-determined values for the non-cooling flow clusters,
inferred from the strong-lensing analyses, are
limited to small ($r \approxlt 100$ kpc) radii. In the following Section,
we shall explore how even these discrepancies may be resolved.

\subsection{Cluster core radii and cooling flows} 
\begin{table*}
\vskip 0.2truein
\begin{center}
\caption{The X-ray masses for the non-cooling flow clusters for a fixed
50 kpc mass core radius}
\vskip 0.2truein
\begin{tabular}{ c c c c c c }
\multicolumn{1}{c}{} &
\multicolumn{1}{c}{} &
\multicolumn{1}{c}{} &
\multicolumn{1}{l}{} &
\multicolumn{1}{l}{} \\
\hline
                      & ~ & $r_{\rm arc}$ &  $M_{\rm arc}$      & $M_{\rm X, r_c50}$      &     RATIO \\
                      & ~ &  (kpc)        &  ($10^{13}$ \Msun)  & ($10^{13}$ \Msun)       & ($M_{\rm arc}$/$M_{\rm X, r_c50}$)   \\
\hline					                              	               
&&&&& \\
NON COOLING FLOWS &&&&& \\
&&&&& \\
Abell 2744(2)         & ~ & 119.6         &  11.36 (9.23)       & $8.71^{+1.35}_{-0.76}$  & $1.30^{+0.13}_{-0.17}$ ($1.06^{+0.10}_{-0.14}$)      \\
Abell 2163            & ~ & 67.7          &  4.29               & $6.09^{+0.60}_{-0.57}$  & $0.70^{+0.08}_{-0.06}$     \\
Abell 2218($\#359$)   & ~ & 79.4          &  5.42               & $4.61^{+0.34}_{-0.44}$  & $1.18^{+0.12}_{-0.09}$         \\
Abell 2218($\#384$)   & ~ & 84.8          &  4.96               & $4.96^{+0.37}_{-0.47}$  & $1.00^{+0.10}_{-0.07}$         \\
Abell 2219 (100 kpc)  & ~ & 100           &  $11 \pm 2$         & $7.76^{+0.84}_{-1.10}$  & $1.42^{+0.53}_{-0.37}$        \\
AC114 (75 kpc)        & ~ & 75.0          &  $4.2 \pm 0.1$      & $5.18^{+0.95}_{-0.77}$  & $0.81^{+0.19}_{-0.14}$ \\
&&&&& \\
INTERMEDIATE &&&&& \\
&&&&& \\
Abell 963             & ~ & 79.8          &  5.85               & $3.78^{+0.52}_{-0.39}$  & $1.55^{+0.18}_{-0.19}$         \\
Abell 1689            & ~ & 183           &  30.7 (27.5)        & $14.7^{+1.8}_{-1.5}$    & $2.09^{+0.24}_{-0.23}$ ($1.87^{+0.21}_{-0.20}$)      \\
&&&&& \\		      
\hline		      
&&&&& \\
\end{tabular}
\end{center}
\parbox {7in}
{Notes: The modified X-ray mass results for the non-cooling flow and intermediate
clusters with the assumption of hydrostatic equilibrium relaxed, and 
a fixed core radius for the mass distributions of 50 kpc adopted.
$M_{\rm arc}$ and $r_{\rm arc}$ are the strong-lensing masses
and radii within which those masses are evaluated. Where possible, we
have used the results from the more sophisticated lensing analyses discussed
in Section 4.3. $M_{\rm X, r_c50}$  values are the X-ray masses within
the same radii (and their formal 90 per cent errors) determined with the
fixed 50 kpc core radius. By relaxing the hydrostatic
assumption and adopting core radii for the mass
distributions in agreement with the mean value determined for the
relaxed, cooling-flow clusters, we can account for the bulk of the
discrepancies between the strong lensing and X-ray masses. }
\end{table*}

      The results on the cluster core radii determined from the X-ray analyses
(Table 4, Column 6) demonstrate a clear segregation between the cooling flow 
and non-cooling flow systems. The mean core radius determined for the 6 
cooling-flow clusters is $\sim 50$ kpc  (with a trend for slightly larger 
core radii in the more X-ray luminous systems). 
For the two intermediate systems the value is 80 kpc, and for
the 5 clear non-cooling flows the value is $\sim 300$ kpc. 
For the cooling-flow clusters for which lensing 
studies provide an independent measure of the mass core radius
(PKS0745-191, MS2137.3-2353 and Abell 2390) excellent agreement between
the X-ray and lensing results is observed.  This result on the core radii,  
together with the agreement of the X-ray and lensing masses for the cooling 
flow clusters (Sections $4.1-4.3, 4.5$) and the alignment of the X-ray and
lensing centroids (Section 4.4), strongly suggests that the assumptions of 
hydrostatic equilibrium and approximate-isothermality in the 
mass-weighted temperature profiles in the cluster cores (Section 3.2),
are valid.  For cooling flow clusters, both the X-ray and gravitational
lensing measurements appear to provide an accurate description of the 
gravitating matter. 

  The results on the offsets of the X-ray and lensing
centres for the non-cooling flow clusters (Section 4.4) showed that in
the central regions of these clusters, the X-ray gas is not in hydrostatic equilibrium with the 
gravitational potentials inferred from the strong lensing data.
The mass distributions in the non-cooling flow
clusters generally appear complex, consisting of two or more large mass 
clumps (Section 4.3), and must be evolving rapidly. 
If the assumptions of hydrostatic equilibrium and spherical symmetry 
used in the X-ray analyses are not valid, and
the central ICM in these clusters has been shocked and disturbed by recent 
(or ongoing) merger events, then the core radii inferred from the X-ray 
data may over-estimate the core radii of the total gravitating matter in
the dominant mass clumps. Such a suggestion is consistent with the results from numerical
simulations (Roettiger \etal 1996) and is in agreement with the results 
for Abell 2219, presented here, for which the X-ray data
suggest a core radius of $\sim 200$ kpc, whereas the detailed 
lensing analysis of Smail \etal (1995a) determines a core radius for the dominant 
mass clump of $46 \pm 20$ kpc, in good agreement with the 
values inferred for the relaxed, cooling flow clusters. 

 We have examined whether inflation of the X-ray core radii, due to 
rapid evolution in the cluster potentials, can provide an 
explanation for the strong-lensing/X-ray mass discrepancies in the 
non-cooling flow clusters. To do this, we have estimated 
the masses within the arc radii that are implied when 
adopting a fixed core radius of 50 kpc (in agreement with the
mean core radius determined for the cooling-flow systems). 
The temperatures measured from the ASCA spectra, which define the
velocity dispersions used to parameterize the mass distributions
(Table 4), were assumed to maintain a reasonable
representation of the virial temperatures of the clusters. 
(The simulations discussed in Section 4.1 suggest this to be reasonable.) 
Note that under the standard assumption of hydrostatic equilibrium in the 
X-ray gas, the use of such small core radii in the deprojection analyses
would imply sharply rising temperature
profiles in the central regions of the non-cooling flow clusters, which
are not, in general,  observed in similar, nearby systems  
\eg Markevitch \& Vikhlinin (1997), Ohashi \etal (1997). See also the 
discussions by Miralda-Escud\'e \& Babul (1995) and  
Waxman \& Miralda-Escud\'e (1995).

  The masses within the arc radii for the non-cooling flow clusters,
calculated using a fixed core radius of 50 kpc and the velocity
dispersions listed in Table 4, are summarized in Table 8. Also
listed are the masses within the same radii inferred from the strong-lensing
data. Where possible, the results from the more detailed 
lensing models discussed in Section 4.3 have been used.  
We see that the effects of over-estimating the gravitational core radii, 
from the X-ray data, can full account for the discrepancies
between the strong-lensing and X-ray masses for the 
non-cooling flow clusters.  
Only for the intermediate cluster, Abell 1689, do such affects 
significantly fail to account for the differences between the mass
measurements. For Abell 1689, it appears that material external to the X-ray 
luminous parts of the cluster, viewed in projection along the line 
of sight to the cluster core, is contributing to the lensing mass
(Teague, Carter \& Gray 1990, Girardi \etal 1997). If this material 
is not virialized and/or the gas density in it is low, then it will not
significantly effect the X-ray emission observed.

\subsection{Cooling flows and the probability of detecting strong lensing} 

\begin{figure}
\centerline{\hspace{2.5cm}\psfig{figure=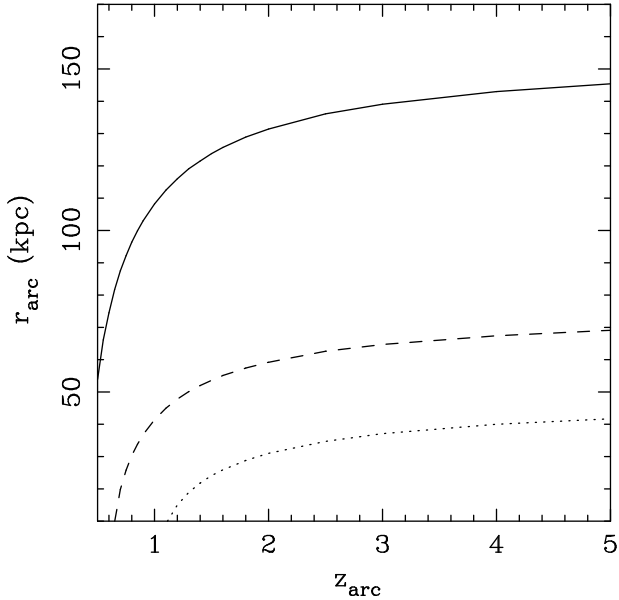,width=0.7\textwidth
,angle=270}}
\caption{The critical radius for strong lensing (putative arc radius) as a 
function of the redshift of potential gravitational arcs in 
three simulated cooling-flow clusters (Section 4.7). The solid 
curve is for a cluster with $\sigma =1000$ \kmps 
(corresponding to $L_{\rm X} \sim  4 \times 10^{45}$ \ergps; {\it c.f.} 
Abell 1835). The dashed curve is for $\sigma =700$ \kmps ($L_{\rm X} \sim  
10^{45}$ \ergps) and the dotted curve for $\sigma =600$ \kmps ($L_{\rm X} 
\sim$ a few $\times 10^{44}$ \ergps). A fixed core radius of 50 kpc
and a cluster redshift, $z=0.3$, are assumed.
}
\end{figure}

  Bartelmann and Steinmetz (1996) discussed how most 
detections of strong lensing in clusters are expected to be made in
clusters of intermediate rather than exceptionally high X-ray luminosity,
and how such detections will be biased (in samples of clusters selected on
the basis of their strong lensing properties) towards systems that are dynamically 
active (\ie systems in the process of virialization or undergoing merger events). 
A significant fraction (6/14) of the 
clusters in our sample have strong cooling flows and exhibit little or no
obvious substructure at optical or X-ray  wavelengths. However, these clusters are 
also amongst the most X-ray luminous and, by implication, most massive systems known. 
At such high X-ray luminosities, even a regular, relaxed cluster presents a 
significant surface area above the critical surface density for strong lensing. 
At lower X-ray luminosities (and masses), however, the area above the 
critical density is reduced, until for $L_{\rm X} < 10^{45}$
\ergps, the probability of detecting strong lensing in a relaxed 
cluster becomes small. This point is illustrated in Fig. 4, where we plot 
the critical lensing radius  (the putative arc radius) 
for three simulated cooling-flow clusters. The clusters have velocity 
dispersions (using the parameterization detailed in Section 3.2) of 
1000, 700 and 600 \kmps, corresponding to X-ray luminosities of 
$\sim 4 \times 10^{45}$, $\sim 1 \times 10^{45}$, and a 
few $\times 10^{44}$ \ergps, respectively. 
A fixed core radius of 50 kpc, a cluster redshift of $z = 0.3$, and 
spherical symmetry are assumed. 
We see that for a luminous ($\sigma = 1000$ \kmps) cooling-flow cluster 
at $z=0.3$, the critical radius varies from $50-150$ kpc for arc 
redshifts, $z_{\rm arc} = 0.5-5.0$.
At luminosities of only a few $\times 10^{44}$ \ergps, however, 
the critical radius is $\approxlt 40$ kpc, and the 
probability of detecting strong lensing is therefore low. 
At such low/intermediate X-ray luminosities, then, the bulk of 
the detections of strong lensing will occur in dynamically active clusters 
where projection effects enhance the lensing signal.
In samples of lensing clusters with low to intermediate X-ray luminosities 
($L_{\rm X,2-10} < 10^{45}$ \ergps), few clusters with 
large cooling flows (in relation to the total cluster luminosity) 
are expected to be found. 

  Finally, we note that a number of other well-known strong-lensing clusters 
have recently undergone, or are awaiting, detailed
ASCA and ROSAT HRI X-ray observations \eg Cl0024+17, Cl0302+17, Abell
2104, Cl2244-02. Comparison with the results obtained for those systems,
using similar analysis techniques to those discussed here, 
will provide an interesting test of the conclusions we have drawn.

\section{Conclusions}

The main conclusions from this work may be summarized as follows: 
\vskip 0.2cm

 (i) Determinations of cluster masses from X-ray and strong gravitational 
lensing data show excellent agreement for cooling-flow clusters (which are
typically dynamically-relaxed systems).  This implies that thermal pressure 
dominates over non-thermal processes in the support of the X-ray gas 
against gravity in these systems, and that the hydrostatic assumption 
used in the X-ray mass determinations is valid. 

 (ii) The mean core radius for the gravitating matter 
in the cooling-flow clusters, determined from the X-ray data, 
is $\sim 50$ kpc (where the core radii have been 
defined using an isothermal parameterization for the total mass profiles). 
Comparisons to measurements of the core radii from detailed 
gravitational-lensing studies show excellent agreement, 
lending further support to the conclusions given in (i).

 (iii) For the non-cooling flow clusters, the masses determined from the strong
lensing data exceed the X-ray values by a factor of $2-4$. 
However, significant offsets between the X-ray and lensing centres are observed,
indicating that the X-ray and strong-lensing data are probing different lines of sight
through the clusters. These offsets, and the generally complex dynamical states 
of the clusters inferred from their X-ray morphologies, lensing data 
and galaxy distributions, suggest that the 
gravitational potentials in the central regions of the non-cooling flow
systems are evolving rapidly, and that the assumption of hydrostatic equilibrium 
involved in the X-ray mass measurements is likely to have broken down. 
The discrepancies between the X-ray and strong lensing mass measurements  
may be reconciled if the dynamical activity has caused the 
X-ray analyses to overestimate the core radii of the dominant mass clumps in 
the clusters. Substructure and line-of-sight alignments of material towards 
the cluster cores may also contribute to the 
discrepancies since they will 
increase the probability of detecting gravitational arcs in the clusters 
and can enhance the lensing masses, without significantly
affecting the X-ray data.

 (iv) Comparisons of the X-ray mass measurements on larger scales 
with measurements from weak gravitational lensing studies 
show excellent agreement for
both the cooling-flow and non-cooling flow systems.

  (v) We have highlighted the importance of accounting for 
cooling flows in X-ray determinations of cluster masses. The 
inappropriate use of simple isothermal models in the analysis of 
X-ray data for clusters with massive cooling flows will result in 
significant underestimates of their X-ray temperatures and masses.

~~
\vskip 1cm

~~

\section*{Acknowledgments}

I thank Andy Fabian for many helpful discussions, Richard Ellis for
comments, and Hans B\"ohringer and Sabine Schindler for kindly providing the ROSAT
HRI image for RXJ1347.5-1145. I also thank the referee, Ian Smail,  
for his detailed and constructive comments regarding the lensing data presented 
in this paper. This research has made use of data obtained 
through the High Energy Astrophysics Science Archive Research Center Online 
Service, provided by the NASA-Goddard Space Flight Center. 
The Digitized Sky Surveys were produced at the Space Telescope Science
Institute under U.S. Government grant NAG W-2166. 
This work was supported by the Royal Society.


\begin{thebibliography}{}
\bibitem{} Allen S.W., Fabian A.C., 1994, MNRAS, 269, 409
\bibitem{} Allen S.W., Fabian A.C., 1997, MNRAS, 286, 583
\bibitem{} Allen S.W., Fabian A.C., Kneib J.-P., 1996a, MNRAS, 279, 615
\bibitem{} Allen S.W., Fabian A.C., Edge A.C., B\"ohringer H., White D.A., 1995, MNRAS, 275, 741
\bibitem{} Allen S.W., Fabian A.C., Edge A.C., Bautz M.W., Furuzawa A., Tawara Y., 1996b, MNRAS, 283, 263 
\bibitem{} Arnaud, K.A., 1996, in Astronomical Data Analysis Software and Systems V, eds. Jacoby G. and Barnes J., ASP Conf. Series volume 101, p17 
\bibitem{} Balucinska-Church M., McCammon D., 1992, ApJ, 400, 699
\bibitem{} Bartelmann M., 1995, A\&A, 299, 11 
\bibitem{} Bartelmann M., Steinmetz M., 1996, MNRAS, 283, 431
\bibitem{} Binney J., Tremaine S., 1987, Galactic Dynamics, Princeton Univ. Press, Princeton
\bibitem{} Buote D.A., Tsai J.C., 1996a, ApJ, 457, 565
\bibitem{} Buote D.A., Tsai J.C., 1996b, ApJ, 458, 27
\bibitem{} Couch W.J., Sharples R.M., 1987, MNRAS, 229, 423
\bibitem{} David L.P., Jones C., Forman W., 1995, ApJ, 445, 578
\bibitem{} David L.P., Harnden F.R., Kearns K.E., Zombeck M.V., 1996, The ROSAT HRI Calibration Report, $ftp://legacy.gsfc.nasa.gov/rosat/doc/hri/hri\_report$
\bibitem{} Dickey J.M., Lockman F.J., 1990, ARA\&A, 28, 215 
\bibitem{} Ebbels T.M.D., Le Borgne J.-F., Pell\'o R., Ellis R.S., Kneib J.-P., Smail I., Sanahuja B., 1996, MNRAS, 281, L75
\bibitem{} Ebeling H., Voges W., B\"ohringer H., Edge A.C., Huchra J.P., Briel U.G., 1996, MNRAS, 281, 799
\bibitem{} Ebeling H., Edge A.C., B\"ohringer H., Allen S.W., Crawford  C.S., Fabian A.C., Voges W., Huchra J.P., 1997, MNRAS, submitted 
\bibitem{} Edge A.C., Stewart G.C., Fabian A.C., 1992, MNRAS, 255, 431
\bibitem{} Eke V.R., Cole S., Frenk C.S., 1996, MNRAS, 282, 263
\bibitem{} Ellis R.S., Allington-Smith J., Smail I., 1991, MNRAS, 249, 184
\bibitem{} Evrard A., 1990, ApJ, 363, 349
\bibitem{} Evrard A., Metzler C.A., Navarro J.F., 1996, ApJ, 469, 494
\bibitem{} Fabian A.C, 1994, A\&AR, 32, 277
\bibitem{} Fabian A.C., Hu E.M., Cowie L.L., Grindlay J., 1981, ApJ, 248, 47
\bibitem{} Fischer P., Tyson J.A., 1997, preprint, astro-ph/9703189
\bibitem{} Fort B., Le Fevre O., Hammer F., Cailloux M., 1992, ApJ, 399, L125
\bibitem{} Fort B., Mellier Y., 1994, A\&AR, 5, 239
\bibitem{} Franx M., Illingworth G.D., Kelson D.D., van Dokkum P.G., Tran K., 1997, preprint, astro-ph/9704090
\bibitem{} Frenk C.S., White S.D.M., Efstathiou G., Davis M., 1990, ApJ, 351, 10
\bibitem{} Fukazawa Y. \etal 1994, PASJ, 46, L55
\bibitem{} Gioia I.M., Maccacaro T., Schild R.E., Wolter A., Stocke J.T., Morris S.L., Henry J.P., 1990, ApJS, 72, 567
\bibitem{} Girardi M., Fadda D., Escalera E., Giuricin G., Mardirossian F., Mezzetti M., 1997, ApJ, in press
\bibitem{} Grossman S.A., Narayan R., 1989, ApJ, 344, 637
\bibitem{} Gudehus D.H., 1989, ApJ, 340, 661
\bibitem{} Henry J.P., Arnaud K.A., 1991, ApJ, 372, 410
\bibitem{} Henry J.P., Lavery R.J., 1987, ApJ, 323, 473
\bibitem{} Johnstone R.M., Fabian A.C., Edge A.C., Thomas P.A., 1992, MNRAS, 255, 431 
\bibitem{} Jones C., Forman W., 1984, ApJ, 276, 38
\bibitem{} Kaastra J.S., Mewe R., 1993, Legacy, 3, HEASARC, NASA
\bibitem{} Kitayama T., Suto Y., 1996, ApJ, 469, 480
\bibitem{} Kneib J.-P., Soucail G., 1995 
\bibitem{} Kneib J.-P., Mellier Y., Fort B., Mathez G., 1993, A\&A, 273, 367
\bibitem{} Kneib J.-P., Mellier Y., Pell\'o R., Miralda-Escud\'e J., Le Borgne J.-F., B\"ohringer H., Picat J.-P., 1995, A\&A, 303,27
\bibitem{} Lavery R.J., Henry J.P., 1988, ApJ, 329, 21L
\bibitem{} Le Borgne J.F., Pell\'o R., Sanahuja B., 1992, A\&AS, 95, 87
\bibitem{} Loeb A., Mao, S., 1994, ApJ, 435, 109L 
\bibitem{} Lucey J.R., 1983, MNRAS, 204, 33 
\bibitem{} Markevitch M., 1997, preprint, astro-ph/9704106
\bibitem{} Markevitch M., Vikhlinin A., 1997, ApJ, 474, 84
\bibitem{} McGlynn T.A., Fabian A.C., 1984, MNRAS, 208, 709
\bibitem{} Mellier Y., Fort B., Kneib J.-P., 1993, ApJ, 407, 33
\bibitem{} Miralda-Escud\'e J., Babul A., 1995, ApJ, 449, 18
\bibitem{} Natarajan P., Kneib J.-P., Smail I., Ellis R.S., 1997, preprint, astro-ph/9706129
\bibitem{} Navarro J.F., Frenk C.S., White S.D.M., 1995, MNRAS, 275, 720
\bibitem{} Nulsen P.E.J., B\"ohringer H., 1995, MNRAS, 274, 1093
\bibitem{} Ohashi T., Honda H., Ezawa H., Kikuchi K., 1997, in Makino F., Mitsuda K., eds., X-ray Imaging and Spectroscopy of Cosmic Hot Plasmas, Universal Academy Press, Tokyo, p. 49
\bibitem{} Oukbir J., Blanchard A., 1997, A\&A, 317, 10
\bibitem{} Oukbir J., Bartlett J.G., Blanchard A., 1997, A\&A, 320, 365
\bibitem{} Pell\'o R., Sanahuja B., Le Borgne J.F., Soucail G., Mellier Y., 1991, ApJ, 366, 405
\bibitem{} Pell\'o R., Le Borgne J.F., Sanahuja B., Mathez G., Fort B., 1992, A\&A, 266, 6
\bibitem{} Peres C.B., Fabian A.C., Edge A.C., Allen S.W., Johnstone R.M., White D.A., 1997, MNRAS, submitted
\bibitem{} Pierre M., Le Borgne J.F., Soucail G., Kneib J.-P., 1996, A\&A, 311, 413
\bibitem{} Porter A.C., Schneider D.P., Hoessel, J.G., 1991, AJ, 101, 1561
\bibitem{} Roettiger K., Burns J.O., Loken C., 1996, ApJ, 473, 651
\bibitem{} Schindler S., 1996, A\&A, 305, 756
\bibitem{} Schindler S., Hattori M., Neumann D.M., B\"ohringer H., 1997, A\&A, 317, 646 
\bibitem{} Smail I., Ellis R.S., Fitchett M.L., Norgaard-Nielsen H.U., Hansen L., Jorgensen H.E., 1991, MNRAS, 252, 19
\bibitem{} Smail I., Hogg D.W., Blandford R., Cohen J.G., Edge A.C., Djorgovski S.G., 1995a, MNRAS, 277, 1 
\bibitem{} Smail I., Couch W.J., Ellis R.S., Sharples R.M., 1995b, ApJ, 440, 501
\bibitem{} Smail I., Ellis R.E., Dressler A., Couch W.J., Oemler A. Sharples R.M., Butcher H., 1997, ApJ, 479, 70
\bibitem{} Soker N., Sarazin C.L., 1990, ApJ, 348, 73
\bibitem{} Soucail G., Mellier Y., Fort B., Mathez G., Cailloux M., 1988, A\&A, 191L, 19
\bibitem{} Squires G., Kaiser N., Babul A., Fahlman G., Woods D., Neumann D.M., B\"ohringer H., 1996, ApJ, 461, 572
\bibitem{} Squires G., Neumann D.M., Kaiser N., Arnaud M., Babul A., B\"ohringer H., Fahlman G., Woods D., 1997a, ApJ, 482, 648
\bibitem{} Squires G., Kaiser N., Fahlman G., Babul A., Woods D., 1997b, ApJ, 469, 73
\bibitem{} Sutherland W., 1988, MNRAS, 234, 159
\bibitem{} Tanaka Y., Inoue H., Holt S.S., 1994, PASJ, 46, L37
\bibitem{} Teague P.F., Carter D., Gray P.M., 1990, ApJS, 72, 715
\bibitem{} Thomas P.A., Fabian A.C., Nulsen P.E.J, 1987, MNRAS, 228, 973
\bibitem{} van Haarlem M.P., Frenk C.S., White S.D.M., 1997, MNRAS, 287, 817 
\bibitem{} Viana P.T.P., Liddle A.R., 1996, MNRAS, 281, 323
\bibitem{} Waxman E., Miralda-Escud\'e J., 1995, ApJ, 451, 451
\bibitem{} White D.A., Fabian A.C., 1995, MNRAS, 273, 72
\bibitem{} White D.A., Jones C., Forman W., 1997, MNRAS, submitted
\bibitem{} White S.D.M., 1992, in Fabian A.C., ed., Clusters and Superclusters of Galaxies, Kluwer, Dordrecht, p. 17
\bibitem{} White S.D.M., Efstathiou G., Frenk C.S., 1993, MNRAS, 262, 1023
\bibitem{} Wu X., Fang L., 1997, ApJ, 483, 62 

\end{thebibliography}
\end{document}